# The Schrödinger Equation Describes a Particular Quantum Geometry

Robert L. Navin*

*Real Time Risk Systems LLC,
104 West 40th St., #1001, New York, NY 10018*

## Abstract

This paper posits the existence of, and finds a candidate for, a variable change that allows quantum mechanics to be interpreted as quantum geometry. The Bohr model of the Hydrogen atom is thought of in terms of an indeterministic electron position and a deterministic metric and the motivation for this paper is simply to try to change variables to have a deterministic position and momentum for the electron and nucleus but with an indeterministic (quantum) metric that reproduces the physics of the Bohr model. This mapping is achieved by allowing the metric in the Hamiltonian to be different to the metric in the space-time distance element and then representing the two metrics with vierbeins and assuming they are canonically conjugate variables. Effectively, the usual *Schrödinger* space-time variables have been re-interpreted as four of the potentially sixteen parameters of the metric tensor vierbein in the distance element while the metric tensor vierbein in the Hamiltonian is an operator expressible as first-order derivatives in these variables or vice versa. I then argue that this reproduces observed quantum physics at the sub-atomic level by demonstrating the energy spectrum of electron orbitals is exactly the same as the usual relativistic Bohr model for the Hydrogen atom in a certain limit. Next, by introducing a single dimensionless running coupling $g$ that shows up in the analogous place as, but in addition to, Planck's constant $\hbar$ in the canonical commutator definition I argue that this allows massive objects to couple to the physical space-time geometry but not massless ones - no matter $g$ value. This claim is based on a fit to the Schwarzschild metric with a few simple assumptions and thus obtaining an *effective* theory of how the quantum geometries at nearby space-time points couple to one another. This demonstrates that this coupling constant is related to Newton's gravitational constant $G$.

*email: robnavin@alumni.caltech.edu

# Table of Contents

# 1. Preamble

The central idea offered here is that the Bohr model of the hydrogen atom might be explained by quantum geometry (QG) merely by noting that the usual model with definite metric and indeterminate electron to center-of-mass distance can be mapped via a well defined change of variable, i.e., not a physically different theory, to a model with definite electron position and indefinite (quantum) metric. This idea is potentially paradoxical because it conflicts with the common presumption that quantum geometry effects, i.e., the space-time foam, can only be observed at the Planck length while at the same time realizing another commonly held idea that all Standard Model physics is the low-energy limit of a quantum geometry/gravity theory.

We would like to choose a minimal mapping to have confidence that this QG reproduces the Bohr model of the atom as closely as possible and that the state-space of the Schrödinger equation plus commutation relations and inner product (*IP*) of the familiar Hilbert space structure are inherited. If possible, one might desire there only remain some global effects that could potentially differentiate the pictures.

# 2. Set-up: Classical Hamiltonian Mechanics

I start at the Hamiltonian formulation of the dynamics of a (classical) massive point particle in 3+1 Minkowski space-time in Cartesian co-ordinates, and natural units, $c = \hbar = 1$, in order to set-up notation, and lay the framework for quantization which we will then implement differently to usual (using summation over repeated indices):

$$H(X^\mu, P_\nu) = \frac{1}{2}\eta^{\mu\nu}P_\mu P_\nu,$$

$$dS^2 = \eta_{\mu\nu}dX^\mu dX^\nu,$$

$$\eta = (\eta_{\mu\nu}), \quad \eta^{-1} = (\eta^{\mu\nu}),$$

$$\eta = \eta^{-1} = diag(1,-1,-1,-1).$$

The equations of motion, including the metric are written in their most general form by introducing a quantity $\lambda$ to parameterize paths,

$$X^\mu = X^\mu(\lambda),$$
$$P_\mu = P_\mu(\lambda),$$

which enables the Hamiltonian and the metric specification to be written as:

$$H(X^\mu, P_\nu) = \frac{1}{2}\eta^{\mu\nu}P_\mu P_\nu, \quad \left(\frac{dS}{d\lambda}\right)^2 = \eta_{\mu\nu}\frac{dX^\mu}{d\lambda}\frac{dX^\nu}{d\lambda}.$$

The equations of motion then, in this most general case, are:

$$\frac{dX^\mu}{d\lambda} = \frac{\partial H}{\partial P_\mu} = \eta^{\mu\nu}P_\nu, \quad \frac{dP_\mu}{d\lambda} = -\frac{\partial H}{\partial X^\mu}.$$

Introducing the parameter $\lambda$ allows us to handle massless and massive particles.

The Poisson bracket on functions $f = f(X^\mu, P_\nu, t)$ and $g = g(X^\mu, P_\nu, t)$ on phase-space is defined as

$$\{f, g\} = \frac{df}{dX^\mu}\frac{\partial g}{\partial P_\mu} - \frac{df}{dP_\mu}\frac{\partial g}{\partial X^\mu}$$

and this can be used to reformulate the equations of motion but a principal observation is that constants of motion can be found because

$$\frac{df}{dt} = \frac{\partial f}{\partial t} + \{f, H\}.$$

The Hamiltonian itself turns out to be a conserved quantity in the case of a Minkowski space-time in pseudo-Cartesian co-ordinates, i.e.

$$\eta_{\mu\nu} = diag(1, -1, -1, -1),$$

and is either 0 or a constant $m^2/2$ for a massless and massive particle respectively.

In curved space-times and external electromagnetic fields the Hamiltonian may not turn out to be conserved, as the energy of the system may receive extra input via the gravitational and electromagnetic fields. Other quantities, expressed as functions on phase-space, besides the Hamiltonian may then turn out to be conserved quantities and aid with finding the solutions of the equations.

Let us solve for the simplest case of a massive particle in flat space-time and no external fields. For non-zero mass we can make the arbitrary choice of $\lambda = S$, the rest-frame time of the massive particle, to write, for observer at $X^\mu = (X^0, \vec{0})$, particle at $X^\mu = X^\mu(S)$ with momentum $P_\mu = P_\mu(S)$. The Hamiltonian itself turns out to be a conserved quantity and is $m^2/2$ using the invariant rest-frame mass $m$ of the particle. The solutions describe a particle that travels at a constant velocity, say $V_X$ along the x-axis,

$$\dot{X}^\mu = \frac{dX^\mu}{dS}, \quad P_\nu = ((\gamma m, -\gamma m V_X, 0, 0)),$$
$$X^\mu(S) = (\gamma(S + V_X X), \gamma(X + V_X S), Y, Z), \quad \dot{X}^\mu = (\gamma, V_X \gamma, 0, 0),$$
$$\gamma = \frac{1}{\sqrt{1 - V_X^2}},$$

and obviously includes being at rest at the origin

$$X^\mu = (S, \bar{0}),$$
$$P_\nu = (m, \underline{0}).$$

## 3. Quantum Mechanics: Canonical Variables, Commutators and Vierbeins

We can now review relativistic quantization and let us stick to flat Minkowski space. Typically the Poisson bracket of the phase-space co-ordinates of the Hamiltonian formalism

$$\{X^\mu, P_\nu\} = \delta^\mu_\nu,$$

is replaced with a commutator in the following manner:

$$[X^\mu, P_\nu] = -i\delta^\mu_\nu,$$
$$\Rightarrow P_\mu = i\frac{\partial}{\partial X^\mu};$$
$$dS^2 = \eta_{\mu\nu} dX^\mu dX^\nu,$$
$$P^\mu = i\eta^{\mu\nu}\frac{\partial}{\partial X^\nu},$$
$$\Rightarrow [X^\mu, P^\nu] = -i\eta^{\mu\nu}.$$

One needs to be careful about operator orderings and definitions when there is spatial dependence in the metric, even in flat space-times, for instance, in polar co-ordinates. But the formalism describes a quantized particle in a possibly curved space-time with possibly external fields as well. The equations of motion become a wave equation with the value of the metric as a separate input. In flat space-time, this quantization leads directly to the Klein–Gordon (KG) equation for a particle of mass $m$ and the non-relativistic limit describes a spin-less scalar obeying the Schrödinger equation.

Introduce a vierbein of the form

$$\eta_{\mu\nu} = \eta_{ij} e^i_\mu e^j_\nu, \quad \eta_{ij} = diag(1,-1,-1,-1)$$

together with its natural inverse (assuming the determinant is non-zero)

$$\eta^{\mu\nu} = \eta^{ij} e^\mu_i e^\nu_j, \quad \eta^{ij} = diag(1,-1,-1,-1),$$
$$(e^\mu_i) = (e^i_\mu)^{-1},$$

and we can expand the formalism by replacing the Hamiltonian operator with its Dirac "square-root" of the Hamiltonian, using Dirac matrices, $\gamma^i$, the standard 4x4 real matrices satisfying $\{\gamma^i, \gamma^j\} = \eta^{ij}$, and so generally

$$\gamma^\mu = e^\mu_j \gamma^j,$$
$$\{\gamma^\mu, \gamma^\nu\} = \eta^{\mu\nu},$$

which obviously only makes sense for the quantum theory and describes a spin $\frac{1}{2}$ representation of the Lorentz invariance group, *SO(1,3)*, of the relativistic theory:

$$H(X^\mu, P^\nu) = \gamma^\mu P_\mu,$$
$$dS^2 = \eta_{\mu\nu} dX^\mu dX^\nu,$$

and again this has the non-relativistic limit of the Schrödinger equation but this time with an extra two states (with energy degeneracy) corresponding to the two eigen-vectors of the (arbitrarily chosen when there is no external field) z-axis spin operator, eigen-values $\pm \frac{1}{2}$.

This spin 1/2 representation is the relativistic Bohr model of the atom after introducing an electromagnetic potential of form $A^\mu = \left(-\frac{e}{r}, 0, 0, 0\right)$,

$$P_\mu \mapsto P_\mu + eA_\mu,$$
$$A_\mu = \left(-\frac{e}{r}, 0, 0, 0\right).$$

The full relativistic case can be solved analogously to the non-relativistic case in polar co-ordinates with polynomials in these co-ordinates and provides a relativistic Bohr model of the atom and has corrections to the non-relativistic case that have been observed for the Hydrogen atom using spectrometry.

## 4. Alternative Quantization: Quantum Geometry

I look for a quantum geometry completely equivalent to the formalism of relativistic quantum mechanics (and hence well-defined) simply by implementing the Poisson brackets to commutators in a different manner. I use vierbeins and include the critical step and the core of this paper: *I let the metric in the Hamiltonian be different to the metric on the space-time.*

$$\eta_{\mu\nu} = \eta_{ij} e^i_\mu e^j_\nu, \quad \eta_{ij} = diag(1,-1,-1,-1)$$
$$N^{\mu\nu} = \eta^{ij} E^\mu_i E^\nu_j, \quad \eta^{ij} = diag(1,-1,-1,-1)$$
$$H(X^\mu, P^\nu) = \frac{1}{2} N^{\mu\nu} P_\mu P_\nu$$
$$dS^2 = \eta_{\mu\nu} dX^\mu dX^\nu$$
$$\eta = (\eta_{\mu\nu}), \quad N = (N^{\mu\nu}), \quad N \neq \eta^{-1}.$$

In one place I have used the co-variant metric tensor and in the other the contravariant metric tensor simply to make the formalism less messy below.

In the massive case, parameterizing the particle's motion with rest-frame time, $S$, as above, the commutators become:

$$X^\mu(S) = (\gamma(S + V_X X), \gamma(X + V_X S), Y, Z), \quad P_\nu = ((\gamma m, -\gamma m V_X, 0, 0))$$
$$\dot{X}^\mu = \frac{dX^\mu}{dS} = (\gamma, V_X \gamma, 0, 0),$$
$$\gamma = \frac{1}{\sqrt{1 - V_X^2}},$$
$$\left[\Delta S \frac{dX^i}{dS}, P_j\right] = \Delta S \left[e^i_\mu \dot{X}^\mu, E^\nu_j P_\nu\right] = \Delta S \dot{X}^\mu P_\nu \left[e^i_\mu, E^\nu_j\right] = -i\delta^i_j,$$

and, I then assume, this is implemented in a differential representation as

$$\left[e^i_\mu, E^\nu_j\right] \approx -i\delta^\mu_\nu \delta^i_j.$$

The Hamiltonian metric and the space-time metric are now operators but, as usual, one or the other may be chosen to be algebraic.

The pre-factor $\dot{X}^\mu P_\mu$ is a simple deterministic vector/co-variant vector dot product and I presume it is a valid operation and it calibrates the state density of the space on which an *IP* is being implied. This formula augments the quantum geometry equations of motion, below. One could assume it has the value *m* but it later turns out to be the coupling of mass to the space-time and I introduce a dimensionless effective coupling parameter, $g$, but assume it has value $\frac{1}{m\Delta S}$ for now and so with rescaling write:

$$[e^i_\mu, E^\nu_j] = \frac{-ig}{m\Delta S} \delta^\mu_\nu \delta^i_j.$$

This means with arbitrary momentum and position derivative vector

$$\left[\Delta S \frac{dX^i}{dS}, P_j\right] = -ig\delta^i_j,$$

and so clearly $g$ must be 1 for quantum mechanics (QM) to "work" in this local QG, i.e., along any single time-line. However, I retain the parameter in order to later connect this local frame QM/QG to a theory across a lattice of points, i.e., space-time.

I choose to avoid using possibly large values for the co-ordinates *X* and so use the derivative of the position vector and a rest-frame time variable $\Delta S$ from an arbitrary *start* point. Lastly, the inverse mass ensures the corresponding momentum operator does not have mass in it by cancelling with the mass in the time slot of the momentum vector. It is a little awkward.

A derivative representation of the commutator admits two obvious ways to implement the quantum geometry mathematically:

$$E^\mu_i = \frac{ig}{m\Delta S} \frac{\partial}{\partial e^i_\mu}, \quad or \quad e^i_\mu = \frac{-ig}{m\Delta S} \frac{\partial}{\partial E^\mu_i}.$$

I will call these two pictures the *algebraic space-time metric representation* and the *algebraic Hamiltonian metric representation* respectively.

Note the particle (co-variant vector) momentum and parameterized (contravariant vector) position, and their first derivatives to the parameter *S* are simply deterministic numbers in both representations: the exact momentum *and* exact location of the particle in a quantum space-time. The space-time fuzzes up the distances and geometry to ensure the Uncertainty Principle is not violated by knowing the exact momentum and position *magnitudes* in such a way to exactly reproduce the relativistic KG field, and therefore the Schrödinger equation in the non-relativistic limit. Indeed looking for these equations and matching to the non-relativistic Schrödinger equation is equivalent to defining time in space-time by using a *small* atomic clock to measure time and defining distance using a *small* quantum De Broglie wavelength generator to measure distance.

Before writing out the equations of motion, consider the metric as an operator in this picture and the mathematical machinery needed to put this on a solid footing. In space-time we have a (contravariant) vector space (a differentiable manifold, i.e., a collection of overlapping co-ordinate patches that are at least first order differentiable functions) and the mapping to a covariant vector space is an operator. See Figure 1.

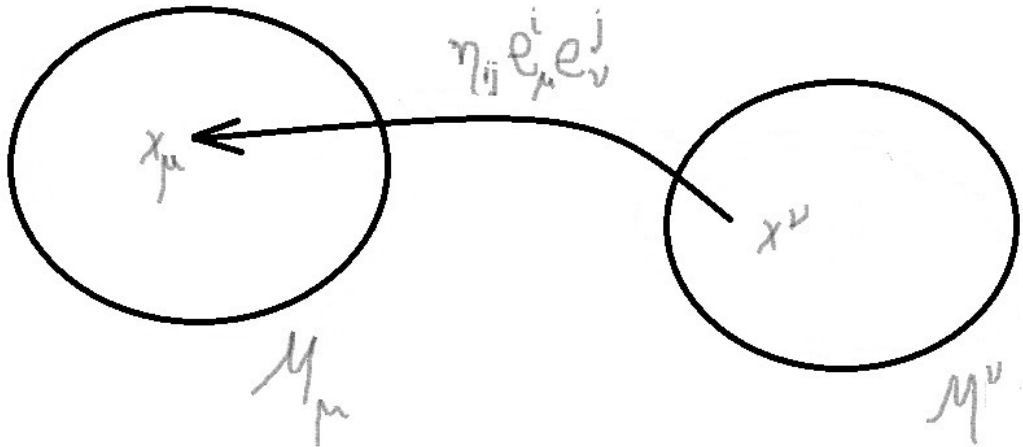

*Figure 1. The metric mapping from manifolds $M^\nu$ to $M_\mu$ is replaced by an operator, leaving $M^\nu$ as a differentiable manifold.*

We can simplify the notation a little by noting the local Lorentz variables usually associated with the vierbein have also been replaced by operators, see Figure 2. :

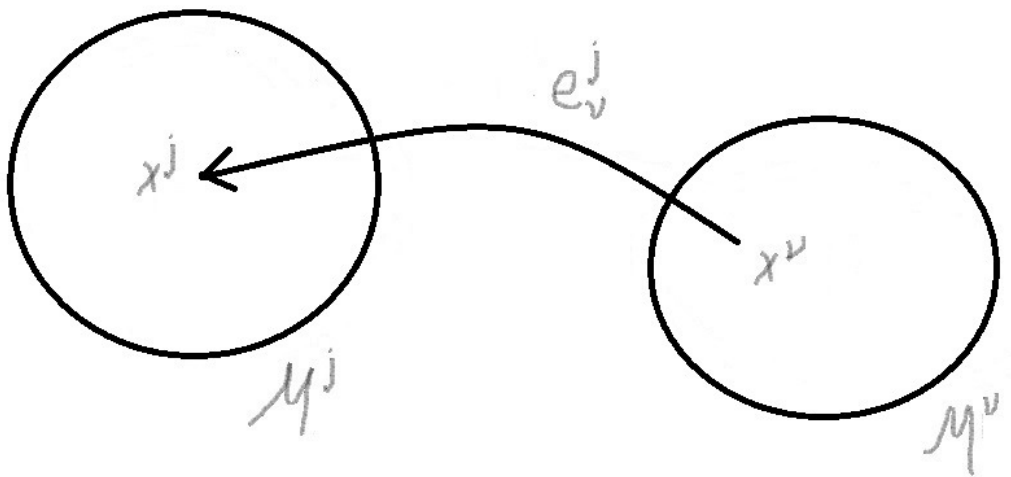

*Figure 2. The vierbein $e^j_\nu$ mapping from manifolds $M^\nu$ to $M^j$ becomes an operator.*

This means that in this picture the position of the particle is a vector, $X^\nu$, populated with deterministic numbers and in general the covariant vector $X_\mu$ and the intermediate "Lorentz" vector, $X^j = e_\nu^j X^\nu$, are only defined as operators on quantum space-time states that may be represented by a wave-function over state variables.

The reverse applies to momentum in the co-ordinates I have chosen; the momentum co-variant vector $P_\mu$ is a set of deterministic numbers although the symmetry of differential geometry surely means these are all arbitrary choices. Then, by choosing one or the other of the two operator pictures for the metric, $E_i^\mu = \frac{ig}{m\Delta S} \frac{\partial}{\partial e_\mu^i}$ or $e_\mu^i = \frac{-ig}{m\Delta S} \frac{\partial}{\partial E_i^\mu}$ we are free to find an algebraic (due to an algebraic and therefore invertible metric) version for one or the other (but not both) of the vectors, $P^\mu$ and $X_\nu$ and the corresponding Hamiltonian and distance metric operators. Essentially one or the other of these two vectors (the covariant position and contravariant momentum) becomes a first order derivative operator in a variable that is the value of the other but they both require expectation values to be taken in the usual way for states on Hilbert space.

In the algebraic space-time metric representation, where the vierbein is algebraic in the space-time metric and a pure first order differential operator on the phase-space, i.e., in the Hamiltonian, and also noting that the vierbein operators are *not* position dependent in the case we consider a flat space-time plus restricting to pseudo-Cartesian co-ordinates to avoid connections for simplicity:

$$E_i^\mu = \frac{ig}{m\Delta S} \frac{\partial}{\partial e_\mu^i}$$

$$\eta_{\mu\nu} = \eta_{ij} e_\mu^i e_\nu^j, \quad N^{\mu\nu} = \eta^{ij} E_i^\mu E_j^\nu,$$

$$H|\psi\rangle = \frac{g^2}{2(m\Delta S)^2} \eta^{ij} P_\mu P_\nu \left( -\frac{\partial^2}{\partial e_\mu^i \partial e_\nu^j} \right) |\psi\rangle$$

$$\Sigma^2 |\psi\rangle = \eta_{ij} e_\mu^i e_\nu^j \frac{dX^\mu}{dS} \frac{dX^\nu}{dS} |\psi\rangle.$$

Note that we have an operator given by an algebraic expression, in the components of the metric operator, $\Sigma^2$, because I chose this particular quantum geometry representation, in the same way that in Schrödinger QM the position operator acts on the wave-function in the spatial wave-function representation (as opposed to the Fourier Transform representation). Then these equations of motion become:

$$H|\psi\rangle = \frac{g^2}{2(m\Delta S)^2}\eta^{ij}P_\mu P_\nu\left(-\frac{\partial^2}{\partial e^i_\mu \partial e^j_\nu}\right)|\psi\rangle = \frac{m^2}{2}|\psi\rangle$$

$$\Sigma^2|\psi\rangle = \eta_{ij}e^i_\mu e^j_\nu \frac{dX^\mu}{dS}\frac{dX^\nu}{dS}|\psi\rangle = \sigma^2|\psi\rangle$$

$$\sigma^2 = 1.$$

The Hamiltonian operator equation reduces to the KG equation on the quantum space-time states as desired.

I want to change variables to identify what are usually understood as time and space variables in the Schrödinger picture of a quantum system. To flesh this out with the simplest concrete example: choose a particle at the origin in pseudo-Cartesian co-ordinates with parameterized position and momentum as follows:

$$\dot{X}^\mu = (1,\overline{0}), \quad P_\mu = (m,\underline{0}),$$

$$\dot{X}^i = e^i_\mu \frac{dX^\mu}{dS} = e^i_0 = \frac{g}{\Delta S}(t,x,y,z),$$

$$\Rightarrow$$

$$e^i_0 = \frac{g}{\Delta S}x^i,$$

i.e., the classically trivial case: a particle at rest at spatial origin. However, the quantum geometry means it is no longer as trivial. I have arbitrarily labeled the algebraic variables in the space-time metric vierbein operator and scaled them by rest-frame elapsed time $\Delta S$ and coupling constant $g$. However, one should not lose sight of the fact that they are four of the 16 Hermitian operators that replace the 16 elements in the classical metric. The phase-space vierbein is replaced with pure Hermitian first-order derivative operators in the same variables. Furthermore, in this case the quantum geometry has been *projected* onto the two vectors, $\dot{X}^\mu$, $P_\mu$, as follows:

$$P_\nu E^\nu_j = P_0 \frac{ig}{m\Delta S}\frac{\partial}{\partial e^i_0} = \frac{ig}{\Delta S}\frac{\partial}{\partial e^i_0} = i\frac{\partial}{\partial x^j} = i\partial_i;$$

$$H|\psi\rangle = \frac{1}{2}\left(-\eta^{ij}\partial^2_{ij}\right)|\psi\rangle = \frac{m^2}{2}|\psi\rangle;$$

$$\Sigma^2|\psi\rangle = \sigma^2|\psi\rangle$$

$$\Rightarrow \sigma^2 = \frac{g^2}{\Delta S^2}(t^2 - x^2 - y^2 - z^2) = 1.$$

We see that the wave-equation for the space-time state is just the KG equation with mass $m$ and the variables which are usually interpreted as co-ordinates in space-time are here some of the values in the space-time metric: in fact, just the ones projected onto the position vector derivative $\dot{X}^\mu = (1,0,0,0)$. The projection means that of the potentially 16 variables in the vierbein we are seeing only four.

Secondly, the constraint $\sigma^2 = 1$ that imposes the space-time metric to be correct, leads to a second *unfamiliar* constraint equation:

$$\frac{\Delta S^2}{g^2} = t^2 - (x^2 + y^2 + z^2).$$

I return to this hyperbolic surface in the future light-cone for each value of $\Delta S$ in detail below.

Lastly, note the non-zero spatial velocity and momentum case leads to a Lorentz transformation of the vierbein operators and hence underlines the co-ordinate invariance of the theory.

$$\dot{X}^\mu = \frac{dX^\mu}{dS} = (\gamma, V_X \gamma, 0, 0), \quad P_\nu = ((\gamma m, -\gamma m V_X, 0, 0)),$$
$$e^i_\mu = \gamma e^i_0 + V_X \gamma e^i_1, \quad E^\nu_j = \gamma m E^\nu_0 - \gamma m V_X E^\nu_1.$$

This highlights the need to have all the components of the vierbeins $e^i_\mu, E^\nu_j$ be canonical conjugates and not just the time-like parts in a given frame of reference (FoR), i.e., $e^i_0, E^0_j$, to maintain Lorentz invariance.

The algebraic Hamiltonian metric representation results in the same equations but they show up in the reverse places between the metric and Hamiltonian constraint equations. This equivalence between the representations is a manifestation of Born reciprocity [1] in this QG.

So, in both representations, the two equations of motion we have are:

$$(-\eta^{ij} \partial^2_{ij}) \psi = m^2 \psi;$$
$$\frac{g^2}{\Delta S^2}(t^2 - x^2 - y^2 - z^2)\psi = \psi.$$

The KG equation replicates the results of the usual quantum mechanics and the second equation is a new surface constraint. Consider this further,

$$\dot{X}^{\mu} = (1, \overline{0}), \quad P_{\mu} = (m, \underline{0}),$$

$$\dot{X}^{i} = e_{\mu}^{i} \frac{dX^{\mu}}{dS} = e_{0}^{i} = \frac{g}{\Delta S} x^{i} = \frac{g}{\Delta S}(t, x, y, z),$$

$$\dot{X}_{i} = \eta_{ij} \dot{X}^{j} = \frac{g}{\Delta S}(t, -x, -y, -z);.$$

meaning that the local Lorentz (contra-variant) vector $\dot{X}^{i}$ derivative is a quantum operator, and furthermore,

$$\dot{X}_{\mu=0} = e_{0}^{i} \dot{X}_{i} = \frac{g^{2}}{\Delta S}(t^{2} - x^{2} - y^{2} - z^{2}) = 1,$$

$$\dot{X}^{i} \dot{X}_{i} = \dot{X}^{\mu} \dot{X}_{\mu} = \frac{g^{2}}{\Delta S}(t^{2} - x^{2} - y^{2} - z^{2}) = 1,$$

$$\Rightarrow \eta_{ij} e_{0}^{i} e_{0}^{j} = \eta_{\mu\nu}\Big|_{\mu=0,\nu=0} = \frac{g^{2}}{\Delta S^{2}}(t^{2} - x^{2} - y^{2} - z^{2}) = 1.$$

Clearly this surface constraint ensures that the space-time metric has the value *1* in the time-time entry. However, it also outlines that the four parameters are essentially three independent co-ordinates on a hyperbolic space-like surface and this form allows us to identify $\Delta S$ as rest-frame time and as it increases a space-like surface is mapped out for each constant $\Delta S$ value. This is, in fact, addressing the difficult issue of time in quantum geometry and is identifying, as special, a constant rest-time-value *curved* space-like surface over the internal variables.

Compare this to the ADM decomposition of the metric for the usual 3+1 quantum gravity analysis,

$$dS^{2} = (N^{2} - N^{i} N^{j} h_{ij}) dt^{2} + 2 N^{i} h_{ij} dx^{j} dt + h_{ij} dx^{i} dx^{j},$$

and here the usual argument is that the state-variable is the space-like metric only and the wave-function is then written as a *functional* of the possible value of the 3-metric, *h,* and a *function* of time and the time-like vector *N*.

$$\psi = \psi(t, N^{\mu}; [h_{ij}(x^{j})]).$$

We have actually increased the size of phase-space and included the time-like *N* vector to become a state-variable. Furthermore, the time derivative of the 3-metric is *not* the canonically conjugate momenta as usual in 3+1 quantum gravity that leads to the Wheeler-DeWitt equation: it is the metric in the Hamiltonian. Hence, the QG presented here is very different to the usual QG of 3+1 quantum gravity.

The internal 3D surface constraint outlines how we need to unambiguously define time development coming from the Hamiltonian, i.e., a separation of the Hamiltonian of the form

$$H = i\frac{\partial}{\partial(\Delta S)} + H'$$

$$\psi(\Delta S, [t, x, y, z]_{S_3}) = \exp(-i\Delta S H')\psi([t, x, y, z]_{S_3})$$

where the space-like surface $S_3$ is described by $(t^2 - x^2 - y^2 - z^2) = \frac{\Delta S^2}{g^2}$ for each value of $\Delta S^2$ it is a 3D space-like hyperbolic surface assuming the coupling $g^2$ has value *1* and the Hamiltonian needs to be linearized in the time derivative either by taking an actual square-root of the Hamiltonian or by using the method of Dirac used to linearize the KG equation to produce the Dirac equation (see next section). Then the restriction of the Hamiltonian to this hyperbolic constraint surface $S_3$ is $H'$.

This is how we address the well-known problem of time in quantum gravity [2][3] in the current calculation. This space-like hyperbolic constraint surface is very important for the relativistic causal structure of the theory. At rest-frame time $\Delta S = 0$, an arbitrarily chosen initial rest-frame time, the particle's *position* is spread across the entire light-cone of internal variables and as the rest-frame time increases this position immediately becomes a space-like surface and gets flatter and flatter over increasingly large areas as $\Delta S$ gets larger and larger. The surface tends to flat for $r << \Delta S$. See Figure 3.

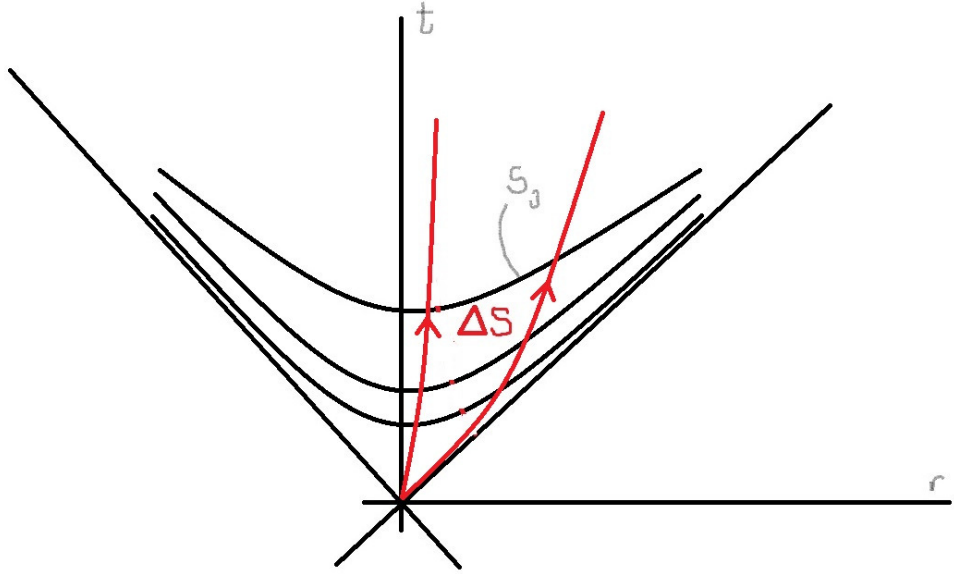

*Figure 3. The hyperbolic space-like surface $S_3$ picked out by quantization assuming the space-time metric and the Hamiltonian metric are canonical conjugates (instead of momentum and position as usual in QM) and parameterized by constant values of $\Delta S$, i.e., rest-frame time for the particle. Lines orthogonal (under the Minkowski pseudo-Cartesian metric) to these surfaces are also drawn, in red, parameterized by $\Delta S$.*

Finally, note, we can introduce a classical electromagnetic potential, $A^\mu = \left(-\dfrac{e}{r}, 0,0,0\right)$, in the following way:

$$P_i \mapsto P_i + eA_i(X^\nu),$$
$$A_i = \left(-\dfrac{e}{r}, 0,0,0\right),$$

by putting the potential into the internal momentum vector and not the space-time momentum vector. This necessarily picks out the algebraic space-time metric representation of quantization as preferred because this allows us to solve in the spatial distance variables in the internal co-ordinates:

$$E_i^\mu = i\dfrac{g}{m\Delta S}\dfrac{\partial}{\partial e_\mu^i}, \quad e_0^i = \dfrac{g}{\Delta S}x^i = \dfrac{g}{\Delta S}(t,x,y,z),$$
$$X^\mu = (S,0,0,0), \quad P_\mu = (m,0,0,0),$$
$$P_i = E_i^\mu P_\mu \mapsto E_i^\mu P_\mu + eA_i(r),$$
$$r^2 = x^2 + y^2 + z^2 = \left(\left(e_0^i\right)^2\Big|_{i=0} - \eta_{ij}e_0^i e_0^j\right)\left(\dfrac{\Delta S}{g}\right)^2$$

and

$$\partial_i \mapsto \partial_i + ieA_i,$$

$$A_i = \left(-\frac{e}{\sqrt{x^2 + y^2 + z^2}}, 0, 0, 0\right).$$

The solution is a relativistic (spin 0 electron) Bohr-model of the Hydrogen atom as a solution over internal space-time variables. Consider the solutions to this equation,

$$\left((\partial + ieA)^2 + m^2\right)\psi = \left(\frac{1}{\sqrt{\eta}}(\partial_i + ieA_i)\sqrt{\eta}\eta^{ij}(\partial_j + ieA_j) + m^2\right)\psi = 0;$$

ignoring the surface constraint will give exactly the same energy eigen-values and therefore energy spectra as the usual model in QM and can be written:

$$\psi_{nlm}(e_0^i) = \exp\left(-\frac{iE_{nlm}t}{\hbar}\right)H_n(r)Y_{nlm}(\theta, \phi).$$

The surface constraint will change this but for the limit of large values of $\Delta S$ we will recover the usual energy spectra. Note that for finite $\Delta S$ the observed discrete energy levels observed will be given by eigen-values of the Laplace operator on the constrained 3-D surface $S_3$,

$$E_h^2 = (\nabla + ieA)^2\Big|_{S_3} = \frac{1}{\sqrt{h}}(\partial_i + ieA_i)\sqrt{h}h^{ij}(\partial_j + ieA_j)$$

with metric $h_{ij}$ with determinant $h$. This metric is given by using the constraint with $\Delta S$ held constant to get

$$ds^2 = dt^2\Big|_{surface} - (dr^2 + r^2d\theta^2 + r^2\sin^2\theta d\phi^2);$$

$$t^2(r) = r^2 + \frac{\Delta S^2}{g^2}$$

$$\Rightarrow dt = \frac{r}{t}dr = \left(1 + \frac{\Delta S^2}{r^2 g^2}\right)^{-\frac{1}{2}}dr,$$

which implies

$$h_{ij}dx^i dx^j = ds^2 = -\left(dr^2\left(1+\frac{r^2 g^2}{\Delta S^2}\right)^{-1} + r^2 d\theta^2 + r^2 \sin^2\theta d\phi^2\right).$$

This energy spectrum shows this theory will reproduce well-tested physics correctly in the *stationary* limit of large rest-frame time variable $\Delta S$. Furthermore, $\Delta S$ is a well-defined time measure for the observer and while the energy spectrum eigen-values will be different for small values of $\Delta S$ it represents the concept of time for a rest-frame observer and is related to the time-like variable of the internal QG. It represents an unperturbed Bohr-atom created at rest at rest-frame time $\Delta S = 0$, and being initially spread out over the light cone in internal variables, then eventually as the constrained surface evolves to a space-like flat surface the energy levels settle to the usual relativistic spin 0 Bohr energy levels.

## 5. Dirac Method: Spin ½ Case

We can find a linearized Hamiltonian using two sets of standard 4x4 real Dirac matrices,

$$\{\gamma^i, \gamma^j\} = \eta^{ij}, \quad \eta^{ij} = diag(1,-1,-1,-1),$$
$$\{\gamma_i, \gamma_j\} = \eta_{ij}, \quad \eta_{ij} = diag(1,-1,-1,-1).$$

To get the spin ½ picture, equivalent to that of the spin ½ particle in QM, note that

$$\gamma^\mu = E^\mu_j \gamma^j,$$
$$\gamma_\mu = e^j_\mu \gamma_j,$$

are to be used in the Hamiltonian and the metric operators and therefore introducing spin as well as linearizing the equations. We clearly need to linearize *both* equations to preserve Born reciprocity.

In summary, using the algebraic space-time metric representation:

$$E^\mu_i = i\frac{g}{m\Delta S}\frac{\partial}{\partial e^i_\mu},$$
$$\mathrm{H} = \gamma^j E^\nu_j P_\nu,$$
$$\Sigma = \gamma_i e^i_\mu \dot{X}^\mu,$$

choose a particle at rest at the origin:

$$X^\mu = (S, \bar{0}), \quad P_\mu = (m, \underline{0})$$

and the same special choice of variables to get:

$$\dot{X}^i = e_0^i = e_0^i = \frac{g}{\Delta S}(t, x, y, z),$$

$$\Rightarrow e_\mu^i \frac{dX^\mu}{dS} = \frac{g}{\Delta S}(t, x, y, z);$$

$$P_\mu E_i^\mu = i \frac{g}{\Delta S} \frac{\partial}{\partial e_\mu^i} = i \partial_i,$$

for which the equations of motion

$$H|\psi\rangle = m|\psi\rangle,$$

$$\Sigma^2|\psi\rangle = 1|\psi\rangle,$$

become

$$\gamma^i (i\partial_i)\psi = m\psi,$$

$$\frac{g}{\Delta S}\gamma_i x^i \psi = \psi.$$

and the wave-function $\psi(\eta) = \psi(x)$ is now a four-component vector as usual in the Dirac representation of relativistic quantum mechanics. The surface constraint equation has appeared again and explicitly writing it out using any representation for the Dirac matrices shows it is of the same form as above:

$$\frac{g^2}{\Delta S^2}(t^2 - r^2) = 1.$$

We can add the same electric potential again and get the usual relativistic Bohr model and the low energy limit plus large $\Delta S$, as argued above, presents itself as the original non-relativistic Bohr-Model with two spin states. This is important and was my original objective, to simply change variables and preserve the entire well-established formalism of quantum mechanics but end up with a re-interpretation of it as quantum geometry. We have identified an extra limit of large rest-frame time $\Delta S$ in addition to the non-relativistic limit to recover the correct Bohr atom Energy eigen-values. This *time* is the particle rest-frame time lapsed since a set-up time and therefore the time over which the state is undisturbed.

# 6. The Hyperbolic Surface Constraint and the Problem of Time in Quantum Geometry

In summary this hyperbolic surface constraint is very important to preserve causality and correctly identify *time* in this QG theory:

$$\eta_{ij} e_0^i e_0^j = \frac{g^2}{\Delta S^2}\left(t^2 - x^2 - y^2 - z^2\right) = 1$$
$$\Rightarrow$$
$$\left(e_0^0\right)^2 - \left(e_0^x\right)^2 - \left(e_0^y\right)^2 - \left(e_0^z\right)^2 = 1,$$

means the metric is not degenerate along the time-like path of the at-rest particle. $\Delta S$ is the time as measured by the observer in the FoR that the particle is at rest and will tend to the internal time for large values, i.e., for

$$r^2 \ll \frac{\Delta S^2}{g^2}$$

on this hyperbolic space-like sheet. It is time lapsed while the state is coherent; undisturbed.

The problem of time in quantum geometry [2][3] is well known. This paper presents a solution defined by inheriting the concept of time from the Schrödinger equation and QM because we mapped this well-defined theory to a QG. Indeed, we inherit the *IP*, Hilbert Space and all of the structure of QM into this theory, with the extra twist of being defined on a hyperbolic space-like 3-surface within the flat space and labeled by a constant value of $\Delta S$. Then, as rest-frame time $\Delta S$ evolves the 3-surface gets flatter and the theory converges to look exactly like QM. This QG is confined to a path along the time-line of an at-rest particle. Its initial behavior at $\Delta S$ very small is something like a point and if we consider the stationary solutions of the relativistic Bohr atom, the constraint to the 3-surface initially describes a highly localized system that spreads out with $\Delta S$ to describe the usual Bohr model after a very short time, roughly

$$\Delta S \gg \frac{r_{Bohr}}{c}$$

where the Bohr radius is of order 0.5 Ångstrom. After this time the curvature of the internal three-surface will not be noticed. This is reminiscent of a creation operator for the stationary system.

In conclusion, the *space-time* like variables of QM may be interpreted as the usual picture of space-time in QM or as *internal* variables in a quantum geometry picture and here demonstrated as some of the values of the space-time metric projected onto the position vector derivative with respect to the rest-frame time *S* with the metric in the Hamiltonian being different; an operator and its entries being the conjugate momenta of the metric entries. The Hamiltonian metric and space-time metrics are canonically conjugate variables and this preserves the Born reciprocity idea of QM in the corresponding QG.

Extending this model away from a single time-line, assuming some coupling of the quantum geometries, will produce a full QFT that presumably will be flexible enough to include the full QFT of the Standard Model but also include a framework for, if not a description of, quantum gravity. Below, I attempt a simple *effective* model on space-time that potentially captures a part of the result after space-like translation and linking of quantum geometries at nearby points.

# 7. Identifying the Coupling Constant $g$ as the Gravitational Constant by Fitting to the Spatial Slice of the Schwarzschild Metric

We have identified an internal time co-ordinate *t* together with flat space-like variables *x*, *y,* and *z* that behave like the Minkowski space-time variables of regular relativistic QM. However, we have also picked out a special rest-frame time variable $\Delta S$ that is time-like over the QM variables but is associated with a curved space-like surface for each value of $\Delta S$.

I would like to generalize this QG which is only defined along a single time-like curve in space-time by imposing some rule for parallel transport of the QG across spatially near space-time points in a possibly curved space-time. Basically, we need to know how nearby quantum geometries interact, thinking in terms of a lattice of quantum geometries to represent space-time.

It has been known for some time that a local Poincaré invariance, which is Lorentz invariance plus translation invariance, can be extended out with simple assumptions and produces a theory that satisfies Einstein like field equations. The connection and vierbeins are treated as the potential fields of a (Poincaré) gauge theory of gravity and the only difference to the usual general relativity theory is that it has torsion in the connection [4]. This fact hints that we might be able to extend our model here to at least fit to the Schwarzschild metric around our massive particle, because this picture was essentially derived in something very similar to a local Lorentz invariant picture and we are looking to extend out to include translations for a fully Poincaré invariant picture.

Our quantum geometry presents itself as a *small* system or a tangential system at a specific position and momentum in space-time and phase-space. See Figure 4.

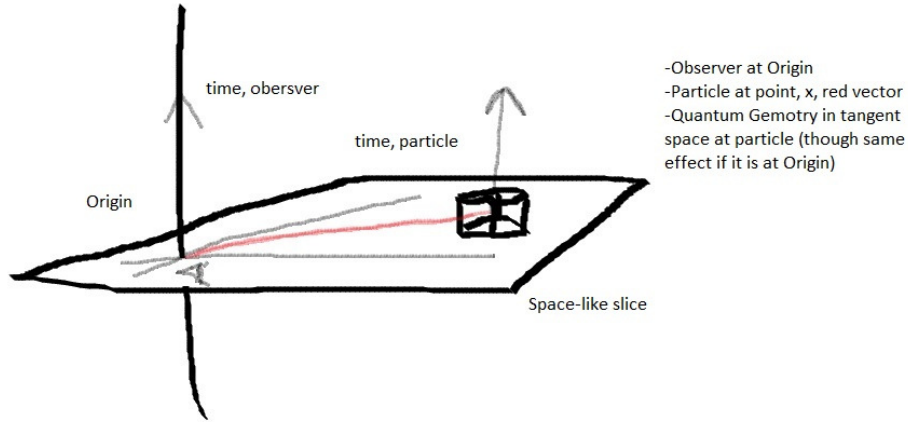

*Figure 4. The tiny quantum clock and measuring device used to probe the 3D space-like surface of space-time and time for the observer.*

The *tiny* quantum geometry can be viewed as located at the particle or at the observer. The metric tensor is really a function of two points on the manifold but they are usually considered infinitesimally far apart. Here, I view them as physically well separated, say the distance of the planet Mercury from the sun!, on a space-time and want to investigate if the *locally flat* (in internal variables) quantum geometry that I have introduced has any discernible effects on this space-time as it evolves per internal co-ordinate time *t*.

The first step is to define parallel transport of the quantum geometry metric onto nearby vectors or equivalently allow the observer to move:

$$e^i_\mu(\delta X^\mu) = e^i_\mu + \Delta^i_\mu.$$

However, I am going to assume the trivial case: the $\Delta^i_\mu$ are all zero and integration up of the internal Lorentz vector introduces co-ordinate shifts only, at least for pure space-like transformations,

$$\Delta^i_\mu = 0$$
$$\Delta S\, \dot{X}^i\Big|_{initial} = e^i_0 \Delta S = g(t, x, y, z)\Big|_{t,x,y,z \in S_3},$$
$$\Rightarrow$$
$$\delta X^\mu = \frac{\Delta S}{g} \dot{X}^i\Big|_{i=\mu} = (t, x, y, z) + X^i_0.$$

This means pure transport of a tiny quantum device to nearby points is almost trivial and not only do we migrate internal time up to the physical world

$$t \approx \Delta S$$

we also migrate up the internal Schrödinger spatial parameters plus their metric and hyperbolic 3-surface constraint as well. This trivial assumption is perhaps the simplest form for a fully quantum geometry field theory derived from QM. Initially, one can think of a lattice across the space-like initial surface with copies of the above quantum geometry at every point but in the limit of the lattice size going to zero the quantum commutators become functional derivatives.

$$e^i_\mu = e^i_\mu(X^\mu),$$
$$E^\mu_i(X^\mu) = \frac{ig}{m\Delta S} \frac{\delta}{\delta e^i_\mu(X^\mu)}.$$

This is the process of "second quantization" in this quantum geometry picture. The functional derivatives are here with respect to spatial surface integrals over 3 spatial co-ordinates,

$$\frac{\delta}{\delta e^i_\mu(X^\mu)} e^j_\nu(Y^\mu) = \delta^3(\overline{X} - \overline{Y}),$$

while we should note this could be with respect to 4D space-time integrals and look like:

$$E^\mu_i(X^\mu) = \frac{ig}{m} \frac{\delta}{\delta e^i_\mu(X^\mu)},$$
$$\frac{\delta}{\delta e^i_\mu(X^\mu)} e^j_\nu(Y^\mu) = \delta^4(X - Y).$$

I will return to these ideas in more detail in the sections below on the vacuum structure.

Now, I will construct an *effective* theory by making the only parameter I have introduced, the dimensionless coupling constant $g$, be a physically spatially dependent quantity: $g = g(X^\mu)$. This will be in addition to the assumption of being able to use internal co-ordinates to measure spatial dependence. We are hence constrained to a surface $S_3$ mapped out by the constrained surface as a constant rest-frame time $\Delta S$ space-like surface:

$$S_3 : t^2 = (x^2 + y^2 + z^2) + \frac{\Delta S}{g^2}.$$

This constrained surface exists in flat 3+1 internal co-ordinates

$$ds^2 = dt^2 - (dx^2 + dy^2 + dz^2)$$

and means for any specific value of $\Delta S$ we can only be located on this hyperbolic surface in the flat Minkowski space of the internal variables as in Figure 5.

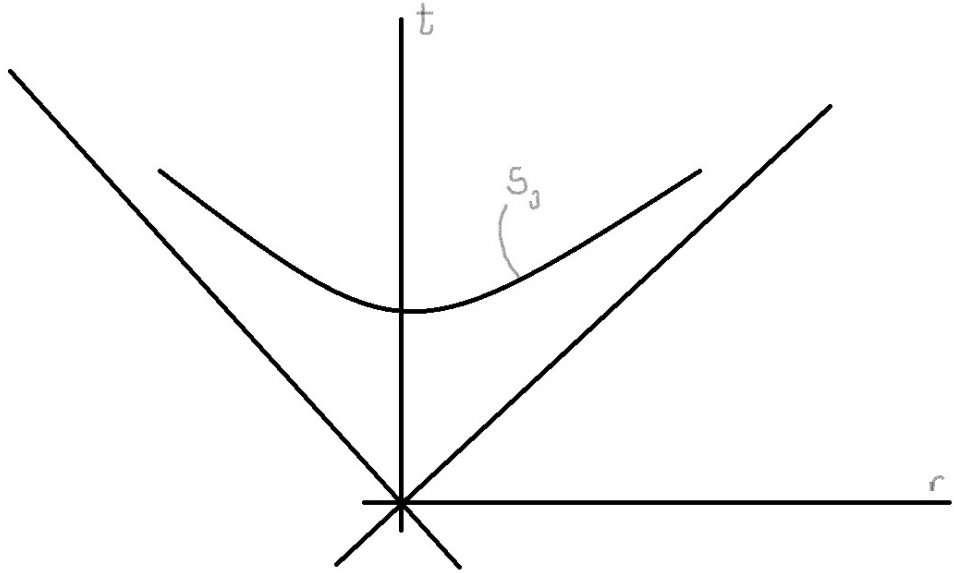

*Figure 5. The tiny quantum clock we use to probe space-time by measuring time and even spatial distances is constrained to a 3D surface $S_3$ of internal space-time and also space-time itself under trivial mapping, given an external rest-frame time $\Delta S$.*

This has already proven to be a space-like slice that corresponds to a constant rest-frame time surface that we now experience in real space-time under the trivial parallel transport above. Using polar co-ordinates having eliminated time by parameterizing transport across the surface using radial coordinate $r$ and assuming the coupling constant $g$ is *constant* over real space-time, the effective metric in this hyperbolic surface is, from above, given by

$$ds^2 = -\left( dr^2 \frac{1}{\left(1 + \frac{r^2 g^2}{\Delta S^2}\right)} + r^2 d\theta^2 + r^2 \sin^2\theta d\phi^2 \right).$$

Returning to the idea that $r$ is very large and describes a distance from some source and presuming the parameter $g^2$ is a space-time $r$ *dependent* coupling constant and should reproduce the spatial part of the Schwarzschild metric, it turns out that there is indeed a form of the coupling constant that does this and the form of it for this effective theory is:

$$g^2 = \frac{\Delta S^2}{G(r)},$$
$$G(r) = -(r + r_+)(r - r_-),$$
$$r_\pm = r_S 2(\sqrt{2} \pm 1),$$

where $r_S$ is the Schwarzschild radius. Note, I have to redefine the 3D surface constraint as

$$t^2(r) - r^2 = G(r) \Rightarrow tdt = \left(r + \frac{G'}{2}\right)dr,$$

$$\therefore$$

$$ds^2 = dt^2\big|_{surface} - \left(dr^2 + r^2 d\theta^2 + r^2 \sin^2\theta d\phi^2\right)$$

$$\mapsto$$

$$ds^2 = -\left(dr^2 \left[1 - \frac{\left(r + \frac{G'}{2}\right)^2}{(r^2 + G(r))}\right] + r^2 d\theta^2 + r^2 \sin^2\theta d\phi^2\right).$$

Substituting the form of $G(r)$ above results in

$$ds^2 = -\left(dr^2 \frac{1}{\left(1 - \frac{r_S}{r}\right)} + r^2 d\theta^2 + r^2 \sin^2\theta d\phi^2\right).$$

This is obviously a quite remarkable result: we have constructed some kind of interpolation formula between quantum mechanics and the Schwarzschild metric. I imposed the constraints of the quantum geometry looking like QM at small distances, and then I looked for an effective long-range theory that looked like General Relativity on space-like slices. Nothing has been discovered here, but the fact you can actually fit both these limits into a relatively simple framework equivalent to the Schrödinger equation is highly provocative.

The coupling constant is imaginary outside a limit a bit larger than the Schwarzschild radius:

$$r > r_-$$

$$g^2 = -\frac{\Delta S^2}{G(r)} < 0$$

$$G(r) = (r + r_+)(r - r_-),$$

$$r_\pm = r_S 2(\sqrt{2} \pm 1).$$

This perhaps is not surprising. Indeed an imaginary $g$ means the quantum geometry is just a regular stochastic theory outside the light-cone, by considering the commutation relations. Indeed, there is a temperature already associated to the metric, with $t$ replaced by $t \to it$ in the Schwarzschild metric, the full 4D Schwarzschild metric becomes a curved surface with positive definite metric and it shows a periodicity in $t$ equal to $2\pi$ divided by a temperature: the Hawking Temperature [5]. The idea being that this Euclidean version of the Schwarzschild metric would be the metric used in thermal Greens functions around black holes. It shows that black holes would be in thermal equilibrium in a box which has walls maintained at this Hawking temperature, and thus consistent with the Hawking radiation idea.

Inside this "thermal" limit the coupling $g$ is everywhere real; the geometry is therefore quantum and even has a well defined value at the origin:

$$r < r_-$$

$$g^2 = \frac{\Delta S^2}{G(r)} > 0$$

$$G(r) = (r_+ + r)(r_- - r),$$

$$r_\pm = r_S 2(\sqrt{2} \pm 1)$$

$$g(0) = \frac{\Delta S}{\sqrt{r_+ r_-}} = \frac{\Delta S}{2r_S}.$$

This clearly relates the coupling constant directly to the Gravitational constant and restoring units,

$$g(0) = \frac{\Delta S c^3}{4Gm}$$

is consistent with the above observations that the massive particle is coupled to the space-time via the hyperbolic surface constraint and the coupling drops out for massless particles which move on null geodesics and correspondingly the 3D constrained surface is the light-cone. Also notice that the implied coefficient of the conjugate momenta are:

$$\frac{g(0)}{m\Delta S} = \frac{c^3}{4Gm^2} \frac{\hbar}{c^2} = \frac{1}{4} \frac{m_P^2}{m^2},$$

including writing it using the Planck mass $m_P = \sqrt{\hbar c / G}$. This means that if we have a particle of mass of order $m_P$ then the value of $\dfrac{g(0)}{m\Delta S}$ is of order 1.

Viewing the constrained surface, by entering the Schwarzschild metric implied effective value for the coupling constant into the constraint formula, shows the surface is space-like at small *r*, shifts to time-like and collapses at the Schwarzschild radius and then immediately outside the Schwarzschild radius the time coordinate is pure imaginary consistent with a Euclidean metric and therefore thermal, See Figure 6.

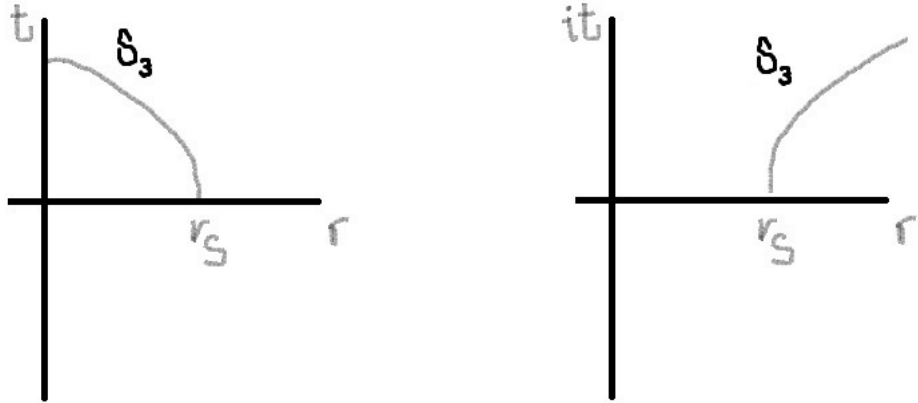

*Figure 6. The shape of constrained surface using the form of coupling constant implied by fitting to the Schwarzschild metric.*

So, the picture emerges that while we have a well-defined quantum geometry at each point we do not have a clear way to parallel transport across the space-like slice to find a full quantum gravity. However, an effective picture is emerging that in the internal space, the theory is quantum in the forward light-cone and thermal outside the light-cone. Then, this system interacts with the nearby points in space-time to excite virtual massive particles via coupling that would emerge from the spatial transport. This *virtual mass cloud* decays away so that for nearby points (a sphere) in the physical space-time we find the thermal region onset is pushed out to be larger than a point, until the decay of the strength of this coupling is such that when the Schwarzschild radius is reached a thermal region outside the light-cone is restored. See Figure 7.

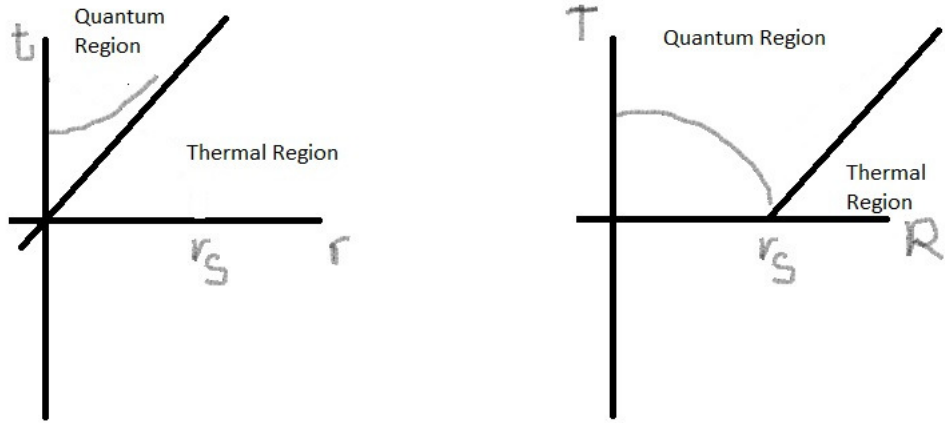

*Figure 7. The internal space-time structure, LHS, where the wave-function of a massive particle is constrained to within the future light-cone, migrates up to physical world, RHS, but a decaying effective coupling strength to nearby points in space-time means those points act as a cloud of massive particles until the decay is such that the thermal region onset is at the Schwarzschild radius.*

Lastly, we should note, every massive point particle, that maps to the Schwarzschild metric at large distances in this effective theory including the electron in the Bohr model, has the potential to be a black hole. However, in the section below outlining an exact fit to the Schwarzschild metric after trying to understand the vacuum, I argue this is not the case for bodies less than Planck mass; the above analysis is probably only valid far outside the Schwarzschild radius for sub-Planck mass elementary particles and down to only around the Schwarzschild radius for Planck mass particles themselves. Below these *cutoffs* there is some kind of cloud of virtual particle interaction giving rise to a renormalized mass that is the observed mass in the Schwarzschild metric.

## 8. Speculation on the Vacuum and Quantum Gravity

This section attempts to dig deeper into the quantum geometry that is equivalent to QM by speculating on the vacuum structure.

Given we have the commutators we can construct typical annihilation and creation operators, Hermitian conjugates of each other, to speculate on the vacuum structure. Sticking to the single point quantum geometry equivalent to a massive particle, outlined above, these operators would look like

$$a^i_\mu = \frac{1}{\sqrt{2}}\left(\eta^{ij}\eta_{\mu\nu}\frac{g}{m\Delta S}\frac{\partial}{\partial e^j_\nu} + e^i_\mu\right),$$

$$a^{i\,\dagger}_\mu = \frac{1}{\sqrt{2}}\left(-\eta^{ij}\eta_{\mu\nu}\frac{g}{m\Delta S}\frac{\partial}{\partial e^j_\nu} + e^i_\mu\right).$$

The *full* Hamiltonian of this theory will be the sum of the *rest-frame energy (squared)* operator, the Hamiltonian above, and the distance element operator

$$H^{Full}_{\mu\nu} \approx m^2 \eta_{ij} a^i_\mu a^{jT}_\nu = H^2_{\mu\nu} + m^2 \Sigma^2_{\mu\nu}$$

$$\Rightarrow$$

$$H^{Full}_{00} \approx \frac{1}{2}\left\{-\frac{g^2}{\Delta S^2}\left(\frac{\partial^2}{\partial e^{t\,2}_t} - \frac{\partial^2}{\partial e^{x\,2}_x} - \frac{\partial^2}{\partial e^{y\,2}_y} - \frac{\partial^2}{\partial e^{z\,2}_z}\right) + m^2\left(e^{t\,2}_t - e^{x\,2}_x - e^{y\,2}_y - e^{z\,2}_z\right)\right\}$$

$$= \frac{1}{2}\left\{-\left(\partial^2_t - \partial^2_x - \partial^2_y - \partial^2_z\right) + m^2 \frac{g^2}{\Delta S^2}\left(t^2 - x^2 - y^2 - z^2\right)\right\}$$

I write this schematically, these operators might represent the Dirac linearized version of the Hamiltonian

$$H^{Full}_\mu = H_\mu + \Sigma_\mu = -i\frac{g}{\Delta S}\gamma^i \eta_{\mu\nu}\frac{\partial}{\partial e^i_\nu} + m\gamma_i e^i_\mu,$$

$$\Rightarrow H^{Full}_0 = -i\gamma^i \partial_i + m\frac{g}{\Delta S}\gamma_i x^i,$$

or "square roots" of the scalar operator versions, thus allowing the first-order time derivative to be factored out of the Hamiltonian and giving us a regular unitary time-evolution picture with well-defined *IP* over space-like surface, i.e., Hilbert space and so on for the eigen-vectors of the energy and distance element operators. It is important to note here that to ensure the Hamiltonian above is recovered

$$H^2_{Full} = H^2_{Energy} + \Sigma^2$$

without new cross-terms we would need to ensure anti-commutators are zero:

$$\{H_{Energy}, \Sigma\} = 0$$

and this would require expanding the Clifford algebra to 8 dimensions from 4, i.e., ensuring that the Dirac matrices in the distance element and the energy Hamiltonian all anti-commute. I suppress this here and continue schematically.

Now, the vacuum for this *full*-Hamiltonian at one spatial point is writable for the single-point quantum geometry, using flat metrics, $\eta$, and inheriting the 3 spatial dimension *IP* from QM plus treating all sixteen entries in the vierbein as independent variables:

$$|0\rangle \approx N_e \exp\left(-\frac{1}{4}\left(\frac{2m_0 \Delta S}{g}\right) e_\mu^i e_\nu^j \eta_{ij} \eta^{\mu\nu}\right),$$

with normalization as required and identifying the mass parameter *m* as a universal mass parameter $m_0$ which is likely to be of order the Planck mass. This implies that at each QG:

$$\langle 0| e_\mu^i e_\nu^j |0\rangle = \frac{g}{2m_0 \Delta S} \eta_{\mu\nu} \eta^{ij}$$

$$\Rightarrow$$

$$\eta_{ij} \langle 0| e_\mu^i e_\nu^j |0\rangle = 4\frac{g}{2m_0 \Delta S} \eta_{\mu\nu}.$$

We might expect this multiplier to be *unity* so that the Planck mass background corresponds to a flat metric. The corollary is that the vierbein commutator at a point is

$$\left[e_\mu^i, E_j^\nu\right] = -i\frac{1}{2} \delta_\nu^\mu \delta_j^i.$$

This would mean that inside the internal FoR of any point there is a QG where $g = 1$ and then we see that the value of $\Delta S$ is of order Planck time. Now let us postulate that every one of the points across a space-like slice of space-time acts as a source that couples to nearby QGs via spatial coupling terms (which we do not have) in the Hamiltonian and while this may lend itself to some kind of renormalization group analysis, the idea is that there is ultimately an effective spatial curvature as the vacuum is excited and it may be synthetically reproduced by having an *effective* coupling constant:

$$\eta_{ij} \langle \psi | e_\mu^i(X) e_\nu^j(X) | \psi \rangle = g_{\mu\nu}(X),$$
$$\Rightarrow g_{\mu\nu}(X) \approx \frac{2g_{eff}(X)}{m_0 \Delta S} \eta_{\mu\nu},$$

where we understand the *IP* to have now been expanded to include a product of vacuum states for all points in a space-time lattice or, equivalently, a vacuum state with integrals over 3D space-like slices. We have now expressed a plausibility argument for the previously assumed effective coupling ideas. The next section tries to add a single massive state to the vacuum, and gets a more general result for this effective theory but allowing an *exact* fit to the Schwarzschild metric.

Returning to the vacuum, we need to examine the vacuum state across space-time. Now at every point across an entire space-like lattice we get

$$|0\rangle \approx \prod_{S_3} N_e \exp\left(-\frac{1}{4}\frac{2m_0 \Delta S}{g} e^i_\mu e^j_\nu \eta_{ij}\eta^{\mu\nu}\right)$$

$$= N_e \exp\left(-\frac{1}{4}\int_{S_3}\frac{2m_0^4 \Delta S}{g} d^3X e^i_\mu(X) e^j_\nu(X)\eta_{ij}\eta^{\mu\nu}\right)$$

and over space-time the commutators and corresponding functional derivative would be something like

$$[e^i_\mu(\overline{X}), E^\nu_j(\overline{Y})] = -i\frac{2g_{eff}(\overline{x})}{m_0 \Delta S}\delta^\mu_\nu \delta^i_j \delta^3(\overline{X}-\overline{Y}),$$

$$\frac{\delta}{\delta e^i_\mu(\overline{X})}e^j_\nu(\overline{Y}) = \delta^\mu_\nu \delta^i_j \delta^3(\overline{X}-\overline{Y}),$$

where the 3D Dirac delta is schematic and may be understood to be spread out over a Planck length size 3-volume if not the corresponding functional analysis limit. The lattice is understood as spaced out by $\frac{1}{m_0}$ and is something like the Planck length. This is provocative because we could write this using 4D integrals over space-time as

$$|0\rangle \approx N_e \exp\left(-\frac{1}{4}\int_{s-t}\frac{2m_0^4}{g} d^4X e^i_\mu(X) e^j_\nu(X)\eta_{ij}\eta^{\mu\nu}\right),$$

where the integral is now schematically over entire space-time. The commutator is now:

$$[e^i_\mu(X), E^\nu_j(Y)] = -i\frac{2g}{m_0}\delta^\mu_\nu \delta^i_j \delta^4(X-Y)$$

$$\frac{\delta}{\delta e^i_\mu(X)}e^j_\nu(Y) = \delta^\mu_\nu \delta^i_j \delta^4(X-Y).$$

The functional calculus is now defined using integrals over 4 space-time dimensions and the vacuum includes an integral over the entire space-time Manifold. There is no concept of time for the vacuum itself. The Hamiltonian introduces the hyperbolic surface constraint at each point that holds only for eigen-functions of the distance element operator and energy operator separately once at least one particle is introduced and induces the 3D *IP* of QM and the concept of time as a low energy approximation. The vacuum is represented by an eigen-function of the full Hamiltonian operator and not necessarily an eigen-function of the two operators and therefore does not necessarily inherit the Hilbert space of QM. This might be remedied by allowing this generalization of the *IP* to an integral over the 4 dimensions of space-time. The full Hamiltonian and the vacuum display the ambiguity of what time means just like the Wheeler-DeWitt equation of 3+1 quantum gravity [2][3].

## 9. Speculation on Single Phonon-type Excitation of the Vacuum: Exact Schwarzschild Metric Fit and a Framework to Approach the Hierarchy Problem

This new *full* Hamiltonian

$$H_{Full} = H_{Energy} + \Sigma$$

$$H_{00}^{Full} \approx \frac{1}{2}\left\{-\frac{g^2}{\Delta S^2}\left(\frac{\partial^2}{\partial e_t^{t2}} - \frac{\partial^2}{\partial e_x^{x2}} - \frac{\partial^2}{\partial e_y^{y2}} - \frac{\partial^2}{\partial e_z^{z2}}\right) + m^2\left(e_t^{t2} - e_x^{x2} - e_y^{y2} - e_z^{z2}\right)\right\}$$

is highly reminiscent of the typical phonon Hamiltonian for a lattice [12]. It contains a lattice vibration *kinetic* energy term, and a *potential* energy term that approximates interaction with nearby lattice points. To speculate further: in this picture massless and massive particles behave like phonons on this lattice of QGs and are almost pure phase-waves that move surprisingly freely across the lattice while gravitons are local lattice excitations of the vacuum and include "twists" and displacement vibration waves corresponding to curvature and gravitational waves. Massless particles couple to the lattice only *very* weakly, if at all, because they are eigen-value zero eigenvectors of both operators, *energy* and *potential*. Massive particles do couple to the lattice, but only weakly, perhaps via higher-order terms that couple across space-time, because they are non-zero eigen-value eigen-vectors of both terms. This coupling causes localized lattice vibrations/dislocations and this is the origin of a renormalized inertial/gravitational mass of value much less than the Planck mass; due to the phonon-type nature of the excitations. Around a massive particle the steady-state solution in the rest-frame is a slightly dislocated lattice experienced as curvature; the Schwarzschild metric and its generalizations to bodies with charge and spin. The phonon-like nature of massless and massive particles allowing very free movement across the lattice also allows arbitrary local Poincaré transforms into different FoRs.

Indeed, taking general ideas about phonons as a guide, excitations across the lattice, i.e., space-time, might be written schematically in the form:

$$\exp(iK_\mu X^\mu)|0\rangle,$$
$$K_\mu K_\nu g^{\mu\nu} = m_R^2,$$
$$m_R \ll m_0$$

and this should be identified as a massless or massive particle, of renormalized mass $m_R^2$, for the eigen-values at the location of a particle of bare mass $m$ if we view the source at the spatial location of a particle as an internal frame excitation of the form

$$\exp(ik_i x^i)|0\rangle$$

where

$$k_i k_j \eta^{ij} = m^2,$$

$$m \ll m_0.$$

The idea would be that momentum vector $k_i$ is the local frame value of a space-time momentum vector $K_\mu$ and the relationship of the two involves some kind of average across coupled lattice points as a source at the particle location:

$$\exp(iK_\mu X^\mu)\Big|_{X^\mu \cong (\Delta S, \vec{0})} \approx \exp\left(i \int_{R_3} k_i e_\mu^i(X) \frac{\Delta S^\mu}{g} d^3 X m_0^3\right)$$

where $R_3 \gg \dfrac{1}{m_0}$ is the region of the coupling via the vacuum, and the integral is over a space-like slice with oriented 3-form volume $\Delta S^\mu d^3 X$. This presumably also allows a renormalized mass $m_R$ and bare mass $m$ to be much less than the order of Planck mass, $m_0$.

Let us try to combine the idea of the vacuum state with a single excitation. It seems time and the Hilbert space are well defined for eigenvectors of the individual energy and potential operators, and the Hamiltonian can be separated out (for the Dirac form most easily) into a time development operator. The vacuum on the other hand seems to have less of a concept of time emphasized writing the vacuum state schematically as

$$|0\rangle \approx N_e \exp\left(-\frac{m_0^4}{4} \int_{s-t} e_\mu^i(X) e_\nu^j(X) \eta_{ij} \eta^{\mu\nu} d^4 X\right)$$

using flat metrics only. Note the integral is well-defined because all exponent index sum values are negative when the vierbein matrices are diagonalized. It also solves the full Hamiltonian at every point. This means there is *no* concept of time for the vacuum, all the manifold directions are equivalent in the vacuum state. Note also that this is the ground state of a Hamiltonian operator describing a system without time development:

$$H_{full}|0\rangle = 0.$$

This absence of time is the typical source of the problem of time in quantum gravity.

Then, adding a single particle that in the time separated version of the (linearized) Hamiltonian looks like

$$\psi \approx \exp\left( iK_0^0 \Delta S m_0 - i \int_{R_3} K_i^\mu e_\mu^i(X) d^3 X m_0^3 \right) |0\rangle$$

which solves the Dirac version of the Hamiltonian and distance element operators

$$\mathrm{H}_{energy} \psi = m \psi$$
$$\Sigma \psi = -m \psi$$
$$\mathrm{H}_{Full} = \mathrm{H}_{energy} + \Sigma.$$

This full Hamilton and its energy part can be separated out at each point in space-time as outlined above,

$$\mathrm{H}_{energy} = i \frac{\partial}{\partial S} + \mathrm{H}'_{energy}$$
$$\Rightarrow$$
$$\mathrm{H}_{Full} = i \frac{\partial}{\partial S} + \mathrm{H}'_{full}$$

to identify a rest-frame time which is in the state vector and schematically written as a solution to the full Hamiltonian as

$$|m\rangle \approx \exp\left( i \int_{s-t} K_i^\mu(X) e_\mu^i(X) d^4 X m_0^4 \right) |0\rangle$$

and then attempting to take the line element expectation value noting we now have no concept of rest-frame time with the functionals defined on integrals over all four spatial dimensions to get

$$\langle m | e_\mu^i(X) e_\nu^j(X) \eta_{ij} | m \rangle = \eta_{\mu\nu} - K_\mu^i K_\nu^j \eta_{ij},$$

An aside to note is that this also works if we make the assumption that the mass state uses the vierbein canonical conjugate:

$$|m\rangle \approx \exp\left( i \int_{s-t} K_\mu^i(X) E_i^\mu(X) d^4 X m_0^4 \right) |0\rangle$$

because this is now a vierbein shift operator.

We then presume that the only values in the $K$-matrix are $K_0^0$ and $K_r^r$ representing a spherically symmetric gravitational disturbance driven by the particle along the time-line at the spatial origin. Fitting to the Schwarzschild metric, written as a flat part plus perturbation,

$$ds^2 = dt^2 - dr^2 - r^2 d\Omega^2 - \frac{r_S}{r}\left\{dt^2 + dr^2\left(1 - \frac{r_S}{r}\right)^{-1}\right\}$$

we get the sign of the time-time entry in the expectation value for the metric correct:

$$\left(K_0^0\right)^2 = \frac{r_S}{r}$$

but the spatial part then gets the sign incorrect outside the Schwarzschild radius:

$$\left(K_r^r\right)^2 = -\frac{r_S}{r}\frac{1}{1-\frac{r_S}{r}}.$$

This possibly reflects the fact already noted above that the system is thermal across space-like slices. It would mean that the gravitational excitations that a mass triggers via coupling to the lattice would be of the form

$$\psi \approx \prod_{s-t}\left\{\exp\left(iK_0^0(R)e_T^T\right)\exp\left(-K_R^R(R)e_R^R\right)\right\}|0\rangle.$$

This is closely consistent with the previously derived value for the effective coupling for large $r$ because we could write this, reintroducing the coupling constant, as

$$\left(K_j^j\right)^2 \mapsto \left(K_j^j\right)^2 \frac{g_{eff}}{m\Delta S}$$

$$\Rightarrow$$

$$\frac{g_{eff}}{\Delta S} = \frac{1}{r}, \quad \left(K_0^0\right)^2 = \frac{2m^2}{m_P^2}, \quad \left(K_r^r\right)^2 = -\frac{2m^2}{m_P^2}\frac{1}{1-\frac{r_S}{r}}.$$

We have a combination of decay of lattice dislocation amplitude in the radial space-like direction and an oscillating mass term in the time-direction. This represents the lattice having been dislocated, curved or twisted, by the presence of a massive particle at the origin. The excitation is driven by the fit of the oscillating time-like part to the mass at the origin and this fit is now to the full 4D Schwarzschild metric whereas above we only fitted to the spatial slice specified by $dt = 0$. Note a massless particle has no effect on the surrounding space-time lattice.

Indeed, this fit of a phonon-type gravitational excitation to the internal frame massive particle excitation allows us to understand the way the coupling constant is scaling away from the particle. Further to the hand-waving speculation above that mapped the internal hyperbolic constraint spatial metric to the spatial part of the space-time metric, we are now able to take advantage of the more precise mapping of the internal time coordinate to space-time rest-frame time. This means we should be able to tie together the time-time part of the phonon excitation at small distances to the internal frame excitation of a particle.

It follows that

$$\psi_{QM} \approx \psi_{s-t}\big|_{r \mapsto r_{min}}$$

where the time part of the internal QM type wave-function is given by (ignoring the factors representing the vacuum and other parts of the wave-function for the moment)

$$\psi_{QM} \approx \exp(imt) = \exp\left(i\frac{m\Delta S}{g} e_0^0\right)$$

and for small $r$, down to a minimum, the time-like part of space-time propagating phonon is

$$\psi_{s-t}\big|_{r \mapsto r_{min}} \approx \exp\left(i\sqrt{\frac{2}{m_P r_{min}}} e_0^0\right).$$

The argument is that they must match up, the oscillations driven by the massive source would not be expected to have larger amplitude than the source itself so there is a minimum distance $r_{min}$ where the two match up. In the internal frame $g = 1$ and so

$$m^2 = \frac{2}{m_P \left(r_{min} \Delta S^2\right)}$$

and the idea would be that there is a minimum lattice size of about Planck length and these phonon excitations cause coherent states over many multiples of the lattice size say $f$ where

$$r_{min} \Delta S^2 = \left(\frac{f}{m_P}\right)^3$$

and so the phonon mass would be (absorbing the factor 2 into $f$ for simplicity)

$$m = \frac{m_P}{f^{\frac{3}{2}}}.$$

This means $f$ has various large values for elementary particles as they are generally small fractions of the Planck mass. Note that for particles of mass much less than the Planck mass we have

$$r_S \ll l_P \ll r_{min}$$

because:

$$f^{-\frac{3}{2}} \frac{1}{m_P} \ll \frac{1}{m_P} \ll f \frac{1}{m_P}$$

and means that, not only are elementary particles not black holes because the QG fuzzes them out far above the Schwarzschild radius, but also, they are interacting with the gravitational background over a much larger distance than the Planck distance and the weakness of this interaction is reflected in how large the multiplier $f$ is. For example, for an electron $f \approx 10^{16}$ and for the proton $f \approx 10^{14}$. It means that over a large number $f$ of Planck lengths the particles are interacting and then produce a Schwarzschild metric outside this distance. We cannot say what this interaction exactly looks like but it does not give rise to the Schwarzschild metric inside this sphere of radius $f$ Planck lengths. The higher mass proton interacts more strongly than the electron and is more damped by the interaction with the graviton vacuum, the interaction is more localized as a result i.e., number of Planck lengths $f$ is smaller.

Outside the Schwarzschild radius (and thus definitely outside the Planck length for a particle of less than Planck mass) the radial part of the phonon wave-function is decaying and this reflects the interaction with the background. So, it seems apparent that higher-order coupling terms are producing the effective gravitational mass observed in the Schwarzschild metric and the majority of the contribution to this interaction is occurring out to the distance

$$r_{min} = f \frac{1}{m_P} = \left(\frac{m_P}{m}\right)^{\frac{2}{3}} \frac{1}{m_P}.$$

So, massive elementary particles couple to the vacuum and generate gravitational or graviton interactions to produce a *mass cloud* over this region less than $r_{min}$. Understanding this interaction and why $f$ takes the large values it does would be the Hierarchy problem. However, we have introduced a framework in which this might just make sense because the idea of phonons which display the characteristic in solid-state physics of unexpected lattice decoupling. Massless particles are perhaps *near-perfect* phonons that do not couple, or only do so *very* weakly, resulting in a purer phonon type propagation and $f \to \infty$. Note also that massive point particles would not be expected to be greater than the Planck mass as these would merely be a super-position of Planck mass excitations of the vacuum of the Hamiltonian. Indeed, the massive phonon-like excitations we have introduced are super-positions of gravitational excitations of the vacuum away from the location of the mass. Lastly, note massless particles could either be prepared in a pure single-point state or as a coherent classical wave across space-time with coherent QG excitations at every point, the only difference would be the observer's choice of state preparation.

Given these phonon-type excitations are apparently generally far less massive than the mass associated with the vacuum, it could well be a consequence that any propagating massive excitation would have a spectrum of possible higher energy excitations, i.e., masses with all other quantum numbers the same. Analysis of phonons in solid state physics typically includes a cutoff for lowest energy, longest wave-length phonons, introduced by identifying one side of the crystal with the opposite side, making the longest wavelengths simple fractions of the crystal size. This is the same as introducing a winding number for the mapping of the phase to a circle. This technique is introduced in phonon calculations to simplify things and make the state-space well-defined as opposed to the physical restriction of crystal size. However, there is an apparent physical equivalent here, the winding number of the mapping of the internal space-like hyperbolic slice to the external space-time. It would be the winding number of the $S^2 \mapsto S^2$ mapping at large $r$ which unwinds over the integral over the graviton induced mass cloud around the particle location. It therefore seems plausible that any massive particle will have a spectrum of higher renormalized-mass copies as a phonon energy spectrum has and with all other quantum numbers the same, together with expecting a larger mass gap from first generation (winding number 0) to second generation (winding number 1) than the other mass gaps, due to the complexity of the unwind which presumably occurs within the virtual-mass cloud region just outlined. Mass generations are, of course, observed for the elementary fermions in nature, at least for three generations.

In summary, the lattice of quantum geometries needs some kind of coupling between nearby *space-time* points to induce an effective curved metric and therefore general relativity on the space-time. This coupling will give rise to the fitting to the Schwarzschild metric outlined speculatively above. This fitting indicated an *effective* value for the coupling that is spatial distance in space-time dependent, and indicates the coupling of the local quantum geometries. It does not affect the analysis that the internal value of $g$ is 1 at a massive particle's timeline, although the very idea of the place at which a massive particle is located could be an average property over some small volume, probably greater than the Planck length, that induces a perhaps even naturally massless phonon with a mass. The Schwarzschild metric fit showed that at small spatial distances

$$\frac{g(0)}{m\Delta S} = \frac{1}{4}\frac{m_P^2}{m^2},$$

which for the background being near flat shows that indeed the background mass parameter is of order Planck mass in the internal QG:

$$m_0^2 = \frac{1}{8} m_P^2.$$

However, this particular result should be questioned because the background fit without massive particles would be expected to be a flat space-time, not a curved one, at least on average. A similar result is hinted at in the phonon calculation above as the elementary particle mass goes to Planck mass, all three length regimes, Schwarzschild radius, Planck length and the radius over which the gravitational mass cloud is induced, all converge at the Planck length.

A final flourish of unconstrained hand-waving allows a parallel when looking at the Einstein Field Equations (EFEs). The *Energy* part of the *Full* Hamiltonian above corresponds to the stress-energy density tensor (operator) $T^{\mu\nu}$ while the line element corresponds to the trace-reversed metric curvature tensor (operator) $G^{\mu\nu}$ in the EFE. So, writing the EFE as a quantum operator equation:

$$\{8\pi G T^{\mu\nu} - G^{\mu\nu}\}|\psi\rangle = 0,$$

the above ideas lead to the speculation that the canonical conjugates are the stress-energy tensor $T^{\mu\nu}$ and trace reversed curvature tensor $G^{\mu\nu}$ in the EFE. These ideas hint that independent metric quantization and matter field quantization do not make sense according to the picture presented in this paper; they should be canonical conjugates and the EFE itself would then be a quantum *harmonic oscillator* equation at each point in space-time as an approximation and hence a phonon-like Hamiltonian.

## 10. Correspondence Principle

Having shown we can link to General Relativity, at least to the Schwarzschild metric anyway, in the above limiting manner I need to return to the case of electromagnetic field potentials. I introduced the electric charge field potential over internal variables in order to reproduce all the states of the relativistic Bohr model of the Hydrogen atom but the Correspondence Principle further requires that we get the same potential at physical distances and hence an inverse distance squared electrostatic force in space-time that will result in Keplerian orbits for charged particle (*packets*) at large distances, at least approximately. However, we should note that the large distance limit of quantum mechanics itself, as expressed by Ehrenfest's theorem [11], doesn't quite get classical physics right for the expectation values of variables at large distances for any potential besides the Harmonic oscillator. The phonon idea is the only candidate for the mechanism where nearby quantum geometries can link together and reproduce the field potential in the internal space of the charged particle at large physical distances from the source particle. The virtual coupling photon behaves as a phonon and thus is able to travel almost unimpeded between QG cells of the lattice, reproducing the electromagnetic field of charged particles like the nucleus at spatially large distances from the nucleus itself, i.e., across nearby quantum geometries. The coupling is essentially gravitational or *inertial* but as with phonons the phase-wave is only very weakly coupled to the lattice. Indeed, renormalization of the masses and couplings of gauge theories is likely to be the only residual gravitational effect of coupling to the lattice at low energies.

It was also noted above that the FoR of any moving particle can be used to quantize the geometry and this fact seems to be another aspect of the same property. The phonons display Lorentz and translation invariance on space-time: a full Poincaré invariance.

## 11. Quantum Field Theory and the Renormalization Group

I want to outline a few ideas to motivate the move to quantum field theory (QFT) from this quantum mechanics picture. Let us consider a massless scalar case and the associated creation operator (factors of mass $m$ and rest-frame time $\Delta S$ all drop out for zero $m$),

$$a_i^{\mu\dagger} = \frac{1}{\sqrt{2}}\left(-g\frac{\partial}{\partial e_\mu^i} + e_\mu^i\right).$$

The idea is that this QG at a *single* space-time spatial point needs to be expanded to nearby points. The massless scalar is an eigenvector of both terms in the full Hamiltonian operator with each reducing to the massless KG equation and the light-cone frontier

$$\nabla^2 \psi = 0,$$
$$(t^2 - r^2)\psi = 0.$$

These operators are defined using the local variables of the QG at the space-time point.

In order to expand to QFT, we allow the operators to act on nearby space-time points on a space-like slice only, where representation now is using functional differentiation

$$[e_\mu^i(\bar{x}), E_j^\nu(\bar{y})] = -ig\,\delta_j^i\,\delta_\mu^\nu\,\delta^3(\bar{x}-\bar{y}),$$
$$E_j^\nu(\bar{x}) = ig\frac{\delta}{\delta e_\mu^i(\bar{x})},$$

and using a massless particle moving along the $x$-axis

$$\dot{X}^\mu = (1,1,0,0), \quad P_\nu = (E_0,-E_0,0,0),$$

We are going to obtain two transverse modes once the solutions to the wave equation are restricted to the light-cone frontier. We get the QFT operators of a massless spin 1 particle:

$$\phi^i(\bar{x}) = \left(e_0^i(\bar{x}) + e_1^i(\bar{x})\right)$$
$$\pi^i(\bar{x}) = \left(E_i^0(\bar{x}) - E_1^i(\bar{x})\right)$$
$$i = 2,3$$

Clearly the natural parallel transport of the massless scalar produced a spin-1 representation and given the original particle position and momentum were specified as vectors this is not surprising. The idea is then that massless particles do not interact gravitationally and do not affect space-time but map out a flat Minkowski space-time. We are allowing the spatial part of the wave operator to transport freely across nearby space-time points' QGs. Somewhat hand-wavingly, but in order to "restore" Lorentz invariance, the Energy part of the Hamiltonian requires the extra terms of the del-squared operator to be added and introduces some of the nearby QG interactions we are speculating about:

$$H = \frac{\partial^2}{\partial t^2} + \frac{1}{2}\sum_{i=1,2}\left[\pi^i(\bar{x})\right]^2 \mapsto \frac{\partial^2}{\partial t^2} + \frac{1}{2}\sum_{i=1,2}\left[\pi^i(\bar{x})\right]^2 - \left[\frac{\partial \phi^i(\bar{x})}{\partial x^{\bar{\mu}}}\right]^2,$$
$$\bar{x} = (x^{\bar{\mu}}), \quad \bar{\mu} = 1,2,3.$$

Now returning to the full Hamiltonian we can speculate that the massive-phonons will get excited and couple only via higher-order terms than the mass operator and the surface constraint. This even leads us to a possible re-interpretation of $\lambda\phi^4$ QFT and the renormalization group. Mass renormalization is not needed unless $\lambda > 0$.

The full Hamiltonian is then

$$\mathrm{H}_{Full} = \mathrm{H} + \frac{1}{g^2} f(e_\mu^i),$$

and phonons, i.e. massless and massive particles, get coupled via higher order terms, i.e.,

$$f(e_\mu^i) \approx m^2 g^2 (e_\mu^i)^2 + \lambda(e_\mu^i)^4 \cdots$$

and if we restrict to a schematic of the full higher order interactions of the spin-1 case we can look at spin 0, $\lambda\phi^4$ QFT. Renormalization allows a finite interaction interpretation of the Feyman perturbation expansion of the interacting theory only when the mass and coupling are allowed to be energy/momentum scale *p* dependent and a distance cutoff $\Lambda$, is introduced:

$$m^2 g^2 = m^2 g^2(p\Lambda),$$
$$\lambda = \lambda(p\Lambda).$$

However, this immediately looks reminiscent of the cutoff I introduced in my effective theory to map the single massive QG excitation to the Schwarzschild metric. The idea is that QFT is an approximation to a full quantum gravity theory and the renormalized mass reflects the distance effects of the phonons, i.e., particles, coupling to gravitons and the respective coupling/screening associated. Note the momentum integrals are over external space-time degrees of freedom in the QG.

This contrasts starkly to the currently accepted idea that gravitational effects are ignored in the QFT formulation of particle physics and renormalization is only dealing with inertial mass; *not* gravitational mass. However, we are developing a picture unifying the renormalized inertial mass of QFT with the gravitational mass of a quantum gravity via gravitons that are a coupling of nearby quantum geometries and that allow higher-order interactions of particles that are themselves low-mass phonon-type excitations of the space-time lattice. A natural scale cutoff just reflects the effective distance that these effects can propagate.

A Renormalization group flow analysis intuitively similar to that of statistical systems will likely play a role to implement these ideas. It encourages one to make a connection between the above effective theory linking the mapped Schrödinger equation to the Schwarzschild metric to a renormalization group analysis of statistical lattice systems near their (second order phase transition) critical point temperature. The general idea would be to look for stable/limiting properties of the theory as the lattice on space-time is coarsened following the ideas of Ken Wilson, e.g., see [13].

In summary, a picture that emerges appears to be that massless particles are phonon-like excitations of the space-time represented by a lattice of QGs that do not couple or *very* weakly couple via gravitation because they are 0-eigenvalue eigenvectors of the energy and distance element operators separately of the full Hamiltonian. They have inertial energy/momentum but not gravitational mass. Then, massive particles are also eigenvectors of the two parts of the full Hamiltonian, but of non-zero eigen-value, i.e., mass *m*. Higher order non-linear terms in the Hamiltonian mean these particles couple to the lattice and these disturbances move across the lattice of QGs; inducing a mass due to local interaction with the lattice and a long-range curvature or disruption of the lattice as well. The higher order effects mean some kind of RG analysis seems relevant to the theory as it scales up from the single QG level to macroscopic space-time. The renormalized mass and coupling from the RG analysis of QFT then appears to be a gravitational effect instead of a pure inertial energy-momentum renormalization. The two terms in the full Hamiltonian, the energy Hamiltonian and the distance element operator may correspond to the stress-energy and space-time curvature tensors and hence imply a parallel with the EFEs. Noting that the two terms are built out of conjugate variables of the QG just the way a quantum harmonic oscillator is constructed or more generally a phonon Hamiltonian for a lattice with implied, if not explicit, higher order interaction terms between lattice sites.

## 12. Observations

I divide the observations into two sections, the first are observations based solely on the mapping of the Schrödinger equation to a quantum geometry; the second set are more speculative as possible extensions of the theory:

*i)* The mathematics presented here are essentially not new. However, with the simple statement that canonically conjugate quantum variables are the vierbein on space-time and the vierbein in the Hamiltonian, instead of position and momentum, we derive a radically new interpretation of all the physics of the Standard Model that is already known. The picture allows particles at deterministic points and a quantum metric to be exactly equivalent to the standard model. This model is a well-defined quantum geometry at each point, inheriting *IP*, Hilbert space and so on from quantum mechanics.

*ii)* This Quantum geometry correctly has the energy spectrum of the relativistic Bohr Model of the Hydrogen atom in a well-defined limit.

*iii)* Born reciprocity is maintained.

*iv)* The fact that I can *force* the Schwarzschild metric into this theory with relatively little effort is very provocative.

More speculative observations based upon trying to extend this point QG theory across space-time and essentially extending the theory to include a full theory of quantum gravity include:

*v)* Some kind of coupling between space-time points will become a quantum gravity theory.

*vi)* We have seen a resolution to the problem of time in quantum gravity by identifying an internal variables space-like surface as special in terms of rest-frame time $\Delta S$ and relating it to the (now *internal*) time co-ordinate *t* that we are familiar with in the KG equation.

*vii)* What metric is mapped out by this process? We have something that looks like $ds^2 = \eta_{00} dt^2 + \cdots$ but we do not really know anything about the space-like part. However, simple ideas for generalizing to transport across a space-like surface can be easily fitted to the Schwarzschild metric.

*viii)* One thing appears definite: quantizing gravity in addition to the Standard Model will not make sense. It would be double counting quantum geometry.

*ix)* Combining the canonically conjugate operators into creation and annihilation operators seems to indicate a new theory with a sum of the usual Hamiltonian and the space-time distance element for its Hamiltonian. A parallel to phonons *x)* on a lattice as excitations of this theory is suggested. These are presumably low-mass or massless elementary particles. Gravitational interactions correspond to twists and vibrations on the lattice itself, i.e., not phonons.

*x)* A hierarchy of masses for particles of any given set of quantum numbers and a non-zero mass seems an almost inevitable conclusion while those without masses would likely not have such a hierarchy. This bears a resemblance to nature.

*xi)* The gravitons of this theory will only interact via higher-order terms in the Full Hamiltonian and they seem to lend themselves to description in a Renormalization Group framework. It might even be related to the renormalization group observed in the Standard Model.

*xii)* There seems to be a hint that the vacuum of this theory might have no concept of *time* with every point in space-time being a quantum geometry but the Hamiltonian has a *symmetry breaking* potential term that simultaneously defines the distance element metric and singles out, at least for massive elementary particles, a timeline that includes the concept of time itself together with the usual spatial 3D Inner Product of quantum mechanics and its Hilbert space, and therefore makes a lattice of quantum geometries over a space-like 3D surface the root building blocks of the physics for these particles, at low energies at least.

*xiii)* It seems plausible that the Bohr model of the atom is *not* describing a *"single"* quantum geometry as these are presumably at the Planck length, but it is some kind of long-range quantum coherence effect, i.e., phonon like, that occurs across many minimum lattice (Planck) lengths of the theory.

## 13. Direction of Future Work

To prove any usefulness for these ideas one would like to calculate something different to the Standard Model: either predict a measurable difference to it or at the very least be able to more easily calculate some quantity we already know to be true of the Standard Model. This should guide future work, my ideas include:

i) *Derive* the form of the effective coupling constant for a massive particle instead of *fitting* to the Schwarzschild metric; describe the coupling of gravitons to space-time and therefore quantum gravity theory and not just quantum geometry at a point. The Renormalization Group could be a productive avenue and the ideas of the RG applied to statistical physics seem promising where various power law properties of observables can be derived that depend only weakly on the specific details of system interactions.

ii) Alternatively, a deeper fit to the full Einstein Field Equations might be productive and thus derive their equivalent formulation in terms of the above *Full* Hamiltonian interaction terms.

- iii) Incorporate some QFT results. See how QFT and quantum gravity separate out of the full quantum gravity theory.
- iv) Develop the structure of the vacuum further.
- v) Perhaps global effects may prove to differentiate this QG from QM. Because I locally forced this theory to reproduce QM, finding a difference to the Standard Model might be contained in global effects, e.g., winding numbers of the above mappings. It might quantize some parameters immediately along the lines of Dirac's quantization of the electric charge being explained by the existence of one magnetic monopole. This might give rise to physical topological defects that correspond to physical phenomena that are different to the Standard Model, we already suspect this may explain the mass generations of fermions.
- vi) Perhaps some result can be calculated more easily in this picture than the Standard Model.

# 14. Conclusion

This paper presents one simple idea: the metric in the Hamiltonian and the metric in the distance element that defines space-time are different and indeed are quantum canonical conjugates to each other. This simple postulate can be shown to replicate the quantum mechanics described by the Standard Model with this *quantum geometry* theory. The introduction of a dimensionless coupling $g$ into the commutators in the same place usually occupied by $\hbar$ can be argued to be related to Newton's gravitational constant, $G$.

# 15. Acknowledgements

About three years ago I asked John Preskill whether the idea of a quantum geometry that reproduces the Bohr atom sounded interesting. Off-the-cuff, he raised the Correspondence Principle as an issue. How do we get general relativity and Keplerian orbits for electrons in a classical electric field out of the picture? Nevertheless, he thought it was an interesting question. This encouraged me to do something to back up the idea and I came up with the toy calculation that I outline in the Appendix. But it is very vague. Three years later, after a lot of thinking in spare moments, brought me to a second (non-relativistic) toy calculation that while provocative was still not very compelling because it required every observer to see a different quantum geometry in order that they all agree on the same properties for the same point-particle, i.e., a non-local theory. Then, I moved to the relativistic case as a last ditch attempt. It was surprising how it fell together over a few months. It even quickly presented a hook to fit to general relativity in a certain limit. The asymptotic match to electronic Keplerian orbits part of Preskill's question was harder and requires the phonon idea to allow massless gauge particles to propagate across the nearby space-time point quantum geometries *effortlessly*. This will be the only way *internal* electromagnetic fields could migrate up to the physical world as electric and magnetic field effects.

The majority of the tools for this work were half-recalled/half-reworked lectures (e.g., [5], [6] jogged my memory) from Part III of the Mathematics Tripos at Cambridge 1987-88 academic year! After the initial draft to jot the ideas down, I then did a cursory literature search and found the claim that there are hints in string theory of the phase-space metric being different to the space-time metric [7], plus this same paper gave a very nice review of the to-me-forgotten idea of Born reciprocity which is highly pertinent to this paper's work. I made a slightly better jab at recalling the problem of time in quantum gravity and looked up a few reviews [2] [3] on this subject as the problem immediately presented itself in the course of this work. The degeneracy of quantum gravity in 2+1 D [8] and below is also a pertinent issue. In addition to the problem of time in quantum gravity, there are various open questions about how a low energy limit of the various candidates for a quantum geometry theory would connect to the world we observe [14]. Expanding the number of variables in the way this paper does is a critical step. Lastly, the extensive work that has been done on a Geometrical version of quantum mechanics since Kibble's observation [9] that the commutators of quantum mechanics have an analogy in the definition of the *IP* in the study of differentiable manifolds seems relevant: review here [10].

I insist on full responsibility for any errors, glaring or otherwise, plus the non-exhaustive literature search I have been able to do and therefore obviously recognize might invalidate this entire exercise and send it up in a puff of irrelevant hot air! That said, any determinative feedback, positive or negative, will be deeply appreciated.

Rob Navin
May 2019

# Appendix

Assuming a connection between Schrödinger's equation and the central limit theorem and its forward Kolmogorov equation, i.e., the heat equation, has provided a central motivation for this work. I lay out here the ideas and back of the envelope calculation using stochastic calculus that provided the motivation for the ideas presented in this paper.

The Wick Rotation ($t \to -it$) of the Schrödinger equation produces the Heat equation, the Forward Kolmogorov equation of the normal process. Now the Central Limit Theorem is often heuristically summarized as saying that all shock distributions with well defined second moment, after repeated application, will eventually converge to a normal distribution. The random walk being the classic example and Brownian motion in gases is another. However, this is not true if the state density of the state variable is not everywhere even, e.g., stock prices are never negative, so they will never be normally distributed while transforming to say daily returns on stock prices lends itself to modeling with a Gaussian distribution. Furthermore, in reverse, we can transform the time and distance variables of the heat equation to obtain many possible non-normally distributed variables in many time-like parameters to give a large class of parabolic differential equations: first order in time-like and second order in space-like variables, including lognormal processes and mean-reverting processes and so on.

So, a subtler reformulation of the central limit theorem would be to say, after observing many iterations of a stochastic process, the variables in which the process looks most *normal* are special, they pick out a special state-variable, *distance*, with state density *constant* and further the growth of the variance of this distribution picks out the *direction* and, analogously to space, the calibration of *time*. The latter statement is clearly intertwined with the second law of thermodynamics.

Practically, small devices with *central limit* tendencies are useful to measure time and space, even on a curved (Riemannian) geometry as well as a flat (Euclidean) space: an hour glass filled with sand will calibrate and measure *time* and a gas' random dispersion will calibrate and measure distance. *Equivalently*, using small devices with *Schrödinger* properties will perform the same tasks: a small clock that consists of many atomic decays to calibrate and measure *time* and the number of de Broglie wavelength's of a quantum particle (such as the electron, but more practically the photon) to calibrate distances.

Lastly, let me draw attention to the idea of changes of variables in stochastic calculus. A fundamental tool for the practitioner of stochastic calculus, applying the theory to real-world processes that may not be normal, is the idea of making a variable change. One can do it on the Kolmogorov equations directly

$$x, t \rightarrow x', t'$$
$$x' = x'(x, t)$$
$$t' = t'(x, t)$$

but this is a potentially very complex and not very intuitive process. Introducing a Wiener process (which describes Brownian motion essentially) for a normally distributed process, such as a gas diffusion density or future stock price distribution density and putting this together with Ito's Lemma for deriving related processes on variables that are an algebraic formula in the underlying state variable, leads to quick and accurate tools to do these variable changes. Non-normal processes such as lognormal and mean-reverting are all obtainable with such variable substitutions.

Given these ideas the Bohr model of the atom might simply require a change of variables to reveal quantum geometry as the root "cause" of quantum mechanics. So, hand-waving, in the formula

$$ds^2 = \eta_{\mu\nu} dX^\mu dX^\nu$$

*either* the co-ordinate distance $dX^\mu$ might be *stochastic* or the metric $\eta_{\mu\nu}$ might be *stochastic*, where by *stochastic* I mean *quantum*.

This has the hallmarks of a *paradox*: it seems to be telling us that the Bohr atom structure is due to the space-time foam of quantum gravity which heuristically one believes is not visible until the Planck Length, i.e., 25 orders of magnitude smaller than the Hydrogen atom!

In summary, perhaps the Schrödinger equation therefore plays a very important role, possibly even in fully relativistic quantum physics, as merely the observer's clock and space calibration device. The fact that a simple Wick Rotation relates the two is also provocative. Further, the intuition I am arguing for is that if we can map the Schrödinger equation *precisely* to a quantum geometry, the actual form of this quantum geometry could provide some illumination to quantum gravity-commonly understood as a quantum geometry itself.

**Toy Calculation**

Let us start with a (charged) scalar particle described by the Schrödinger equation but no other charges around. If we prepare the state to be an approximate eigenvector of position (i.e. a minimum uncertainty packet centered at a certain position, say the origin) then it solves the usual Schrödinger equation

$$i\hbar \frac{\partial}{\partial t} \psi = -\frac{\hbar^2}{2m} \frac{\partial^2}{\partial x^2} \psi$$

This can be solved for a minimum uncertainty packet (see Schiff, Quantum Mechanics, [11]):

$$\psi = \frac{1}{(2\pi)^{\frac{1}{4}}\left(\Delta x + \frac{i\hbar t}{2m\Delta x}\right)^{\frac{1}{2}}} \exp\left[-\frac{x^2}{4(\Delta x)^2 + \frac{2i\hbar t}{m}}\right]$$

and has the pleasant property that its density function looks like a Dirac delta for small variance $\Delta x$.

$$|\psi(x,0)|^2 = \frac{1}{(2\pi)^{\frac{1}{2}}(\Delta x)} \exp\left[-\frac{x^2}{2(\Delta x)^2}\right]$$

Now if we Wick rotate $t \to -it$ then

$$\frac{\partial}{\partial t}\psi - \frac{\hbar}{2m}\frac{\partial^2}{\partial x^2}\psi = 0$$

and this Forward Kolmogorov (i.e. heat) equation can be solved for the minimum uncertainty packet (which is now the Dirac-delta initial-valued Green's function) equivalently,

$$\psi = \frac{1}{(2\pi)^{\frac{1}{4}}\left(\Delta x + \frac{\sigma^2 t}{2\Delta x}\right)^{\frac{1}{2}}} \exp\left[-\frac{x^2}{4(\Delta x)^2 + 2\sigma^2 t}\right]$$

$$\sigma^2 = \frac{\hbar}{m}$$

$$\psi(x,0) = \frac{1}{(2\pi)^{\frac{1}{4}}(\Delta x)^{\frac{1}{2}}} \exp\left[-\frac{x^2}{4(\Delta x)^2}\right].$$

This now also allows us to use the formalism of stochastic calculus. It describes a stochastic process where the change in $x$, $dx$ is described by a Wiener process,

$$dx = \sigma dz$$

where $dz$ is selected from a Normal distribution of variance $dt$, i.e., $\langle dz^2 \rangle = dt$.

The probability distribution of *x* at all times (due to the normal limit theorem) obeys the same forward Kolmogorov (heat) equation at all times

$$\frac{\partial}{\partial t}\psi - \frac{\sigma^2}{2}\frac{\partial^2}{\partial x^2}\psi = 0$$

Now, let us examine the spatial part of the space-time metric, in the usual co-variant formalism of classical geometry:

$$X_i = \eta_{ij} X^j. \qquad [1]$$

Let us assume this describes the flat definite metric of classical physics at all times *t*. The value of $X_i$ is stochastic. This stochasticity can come from the value of $X^j$ OR $X^j$ is fixed and $\eta_{ij}$ is stochastic (and also $X_i$ note).

Now let us prepare a particle in a minimum uncertainty packet at position *X*, *i.e.* $S^2 \gg \Delta x^2$ and note that as time develops we become less certain of the particle's location, it spreads around the point *X* in the usual quantum mechanics picture. We will look at the Wick rotated Schrödinger equation to note that as time *dt* passes,

$$X^i \rightarrow X^i + \sigma dz^i$$

where $dz^i$ are three Weiner processes that obey

$$\langle dz^i dz^j \rangle = \delta^{ij} dt.$$

Now this means that while equation [1] holds initially, an instant later we have

$$\begin{aligned} X_i\big|_{t+dt} &= \eta_{ij}\left(X^j + \sigma dz^j\right) \\ &= \left(\eta_{ij} + d\eta_{ij}\right) X^j \\ \Rightarrow d\eta_{ij} X^j &= \eta_{ik} \sigma dz^k \end{aligned}$$

The most general forward Kolmogorov equation would be:

$$\frac{\partial}{\partial t}\psi\left(\eta_{ij} X^j\right) - \frac{\hbar}{2m}\eta_{im}\frac{\partial^2}{\partial\left(\eta_{ij} X^j\right)\partial\left(\eta_{mn} X^n\right)}\psi\left(\eta_{ij} X^j\right) = 0$$

where we have recognized the stochastic variables as a (generally) non-orthogonal set $\{\eta_{ij} X^j\}$ *i.e.* $\{X_i\}$ the uncertain position of the particle in the covariant co-ordinate space. The particle is located exactly at $(X^i)$ forever with momentum similarly well-defined, presumably momentum zero but its covariant vector position leading to the invariant distance calculation is a quantum variable, induced by the quantum metric. Furthermore, we see this is a subspace of the full metric space, with a projection onto the vector $X_i$.

Wick rotating back gives a *hint* at the quantum geometry theory.

$$i\frac{\partial}{\partial t}\psi(\eta_{ij}X^j) = -\frac{\hbar}{2m}\eta_{im}\frac{\partial^2}{\partial(\eta_{ij}X^j)\partial(\eta_{mn}X^n)}\psi(\eta_{ij}X^j)$$

However, this lacks any intuition about the meaning of these variables and any well specified quantum model that it hints at.